\documentclass[superscriptaddress,groupedaddress,nofootnoteinbib,11pt]{article}
\pdfoutput=1 
\usepackage{jheppub}

\usepackage[utf8x]{inputenc}
\usepackage{mathtools,slashed,mathrsfs}
\usepackage[caption=false]{subfig}
\usepackage{dcolumn}
\usepackage{multirow}
\usepackage{tabularx}
\usepackage{float}
\usepackage{booktabs}
\usepackage{bm}
\usepackage{comment}
\usepackage{setspace}
\usepackage[dvipsnames]{xcolor}
\usepackage[normalem]{ulem} 
\usepackage{enumerate}
\usepackage{siunitx}
\graphicspath{{../Mathematica/}}

\def\AA{\mathcal{A}}

\def\LL{\mathcal{L}}
\def\mat{\begin{pmatrix}}
\def\rix{\end{pmatrix}}
\def\ket{\right \rangle}
\def\bra{\left\langle}
\def\rar{\rightarrow}

\def\dag{\dagger}

\def\({\left(}
\def\){\right)}
\def\[{\left[}
\def\]{\right]}

\def\where{\quad \text{where}\quad}

\def\vec{\mathbf}
\def\vec{\boldsymbol}
\def\MM{\mathcal{M}}
\def\HH{\mathcal{H}}

\def\OO{\mathcal{O}}
\def\TT{\mathcal{T}}

\newcommand\nn{\nonumber}
\newcommand\bea{\begin{eqnarray}}
\newcommand\eea{\end{eqnarray}}

\begin{document}

\title{CMB Spectral Distortions from an Axion-Dark Photon-Photon Interaction}

\date{\today}
\author[a]{Anson Hook,}
\author[a]{Gustavo Marques-Tavares,}
\author[a]{Clayton Ristow}

\affiliation[a]{Maryland Center for Fundamental Physics, University of Maryland, College Park, MD 20742, U.S.A.}

\emailAdd{hook@umd.edu}
\emailAdd{gusmt@umd.edu}
\emailAdd{cristow@umd.edu}

\abstract{

The presence of a plethora of light spin 0 and spin 1 fields is motivated in a number of BSM scenarios, such as the axiverse. The study of the interactions of such light bosonic fields with the Standard Model has focused mostly on interactions involving only one such field, such as the axion ($\phi$) coupling to photons, $\phi F \tilde F$, or the kinetic mixing between photon and the dark photon, $ F F_D$. In this work, we continue the exploration of interactions involving two light BSM fields and the standard model, focusing on the mixed axion-photon-dark-photon interaction $\phi F \tilde F_D$. If either the axion or dark photon are dark matter, we show that this interaction leads to conversion of the CMB photons into a dark sector particle, leading to a distortion in the CMB spectrum. We present the details of these unique distortion signatures and the resulting constraints on the $\phi F \tilde F_D$ coupling. In particular, we find that for a wide range of masses, the constraints from these effect are stronger than on the more widely studied axion-photon coupling.

}

\maketitle

\section{Introduction}
\label{Sec: Intro}

In the past several decades, overwhelming gravitational evidence for the existence of dark matter (DM) has been collected~\cite{Bertone:2004pz,Bergstrom:2000pn}. However, we have yet to observe non-gravitational dark matter interactions with standard model particles. This has led to a wide range of models for particles that could describe dark matter. Of those, a class of very motivated models are ultralight bosonic dark matter models, where dark matter is an ultralight ($m\ll$ eV) scalar or vector field. 

Ultralight bosonic particles, such as axions and dark photons, are well motivated from many points of view.  Extra dimensional theories, such as string theory, typically predict a plethora of light scalars and vectors~\cite{Svrcek:2006yi,Arvanitaki:2009fg,Goodsell:2009xc,Demirtas:2021gsq}.  In fact, already the pioneering work on extra dimensional models by Kaluza and Klein proposed that the electromagnetic gauge symmetry, and thus the photon, could be a consequence of extra-dimensions~\cite{Kaluza:1921tu,Klein:1926tv}. Even without such motivations coming from extra dimensional constructions, both the axion and dark photon are highly motivated candidates for ultralight bosonic dark matter~\cite{Abbott:1982af,Dine:1982ah,Preskill:1982cy,Nelson:2011sf,Arias:2012az,Battaglieri:2017aum}, and, in addition, the axion provides an elegant solution to the strong CP problem~\cite{Peccei:1977np,Peccei:1977hh,Weinberg:1977ma,Wilczek:1977pj}. There are many instances of dark matter consisting of ultralight dark photons produced non-thermally~\cite{Graham:2015rva,Agrawal:2018vin,Bastero-Gil:2018uel,Co:2018lka,Dror:2018pdh,Long:2019lwl}.  One such method for non-thermal production of dark photons is from axions via the coupling in Eq.~\ref{Eq: Interaction}~\cite{longpaper}, providing extra motivation to carefully consider this interaction.

Given the expectation that there might be many light bosons, one of which can play the role of dark matter, one expects interactions between these light particles themselves, as well as interactions between them and the Standard Model.
In this paper, we consider models with two new particles; an axion and a dark photon with a coupling to the standard model of the form
\bea
\label{Eq: Interaction}
\LL\supset \frac{\phi}{2f_a}F^D_{\mu\nu}\tilde F^{\mu\nu} \where \tilde F^{\mu\nu}=\frac{1}{2}\epsilon^{\mu\nu\alpha\beta}F_{\alpha\beta}\; .
\eea
Where $\phi$ is the axion and $F$($F_D$) is the photon (dark photon) field strength tensor.  

Generically, in models with axions and dark photons, interactions of the form $\phi F\tilde F$ and $\phi F\tilde F_D$, as well as kinetic mixing between the photon and dark photon ($FF_D$) would be present, and could be more relevant for detecting the light bosons. However, if one demands that there is a dark charge conjugation symmetry $C_D$ under which the axion and dark photon are odd, these other couplings are absent, or highly suppressed, if there is a small breaking of the symmetry. In  Appx.~\ref{App: Model} we present a simple model that exhibits such a symmetry and show that the coupling given in Eq.~\ref{Eq: Interaction} is the leading one. This coupling has been studied in a variety of scenarios~\cite{Kaneta:2016wvf,Kaneta:2017wfh,Pospelov:2018kdh,Choi:2018mvk,Kalashev:2018bra,Biswas:2019lcp,Choi:2019jwx,Hook:2019hdk,deNiverville:2020qoo,Arias:2020tzl,Hook:2021ous,Carenza:2023qxh}.

Even if one considers models with the $\phi F\tilde F$ interaction, we show the bounds placed from the $\phi F \tilde F_D$ interaction can be stronger. To see this, note bounds placed on a $\phi F\tilde F$ interaction from the Cosmic Microwave Background (CMB) in the presence of an axion dark matter background are placed from polarization measurements~\cite{Fedderke:2019ajk}. Conversely, as we will see in this work, bounds on the interaction in Eq~\ref{Eq: Interaction} are placed from measurements of the CMB frequency spectrum. The CMB spectrum has been measured more precisely than the CMB polarization, implying that the constraints on $\phi F\tilde F_D$ will be stronger than those on $\phi F\tilde F$. Thus, if one considers both interactions to have similar strength, the bounds placed on $\phi F\tilde F_D$ will be stronger. 

The goal of this paper is to investigate how this coupling affects the CMB when either the axion or the dark photon is dark matter. In the early universe, before redshift of $z\sim 1100$, the universe was hot and dense enough that photons, electrons and protons were all in thermal equilibrium with one another. Once the universe cooled to redshift $z=1100$, effectively all electrons were bound to nuclei forming neutral atoms, making the universe transparent to photons, in an era called recombination. As the number of free electrons decreased with the lowering temperature, the mean free path of photons $\lambda_\gamma$ increased. Around the time of recombination ($z \approx 1100$), $\lambda_\gamma\sim H$ and the photons transitioned from being trapped in the electron-baryon plasma to being free streaming. Afterwards, these photons could propagate freely until being detected by CMB experiments. Due to the early thermal equilibrium of these photons, their power spectrum follows that of a blackbody. In the early 1990's, the Cosmic Background Explorer (COBE)~\cite{Mather:1993ij}  satellite equipped with Far Infrared Absolute Spectrophotometer (FIRAS) performed the most accurate measurement of the CMB monopole power spectrum. They found it matched a blackbody with temperature $2.7255$ K to agree to about 1 part in 1,000 to 10,000~\cite{Fixsen:1996nj}, making the CMB monopole power spectrum one of the most precisely measured cosmological observables. Any phenomena that would distort this spectrum is then highly constrained by COBE-FIRAS. Constraints on kinetic mixing~\cite{Caputo:2020bdy} and dark matter interactions~\cite{Slatyer:2018aqg,Zavala:2009mi,Ali-Haimoud:2021lka,Choi:2017kzp,Kunze:2015noa,Berlin:2022hmt} from the COBE-FIRAS data have been placed using these spectral distortions\footnote{Constrains on interactions can also be placed using CMB anisotropies as was done for kinetic mixing in~\cite{Pirvu:2023lch}}. 

The interaction given in Eq~\ref{Eq: Interaction} can cause CMB spectral distortions. In the presence of an axion dark matter background, this interaction allows photons to be converted into dark photons. Likewise, in a dark photon dark matter background, it allows photons to be converted to axions~(see Ref.~\cite{Arias:2020tzl} for an early study of this effect in the resonant regime). These dark sector particles are invisible to us and thus the effect of Eq.~\ref{Eq: Interaction} in both cases is to remove photons from the CMB spectrum. This removal of photons naturally distorts the observed CMB spectrum. In Sec.~\ref{Sec: Distortions}, we will show how the time at which these photons are removed gives rise to different types of distortions and argue that the size of these distortions depends on the probability of removing a photon from the spectrum. In Sec.~\ref{Sec: Probability}, we will compute this probability from the interaction in Eq.~\ref{Eq: Interaction} and in Sec.~\ref{Sec: Computing Distortion} use it to compute the various types of distortions. In Sec.~\ref{Sec: Constraints} we place constraints on the coupling $1/f_a$ by comparing these distortions to the COBE-FIRAS data and briefly comment on the possible shapes of the distortions. We conclude in Sec~\ref{Sec: Conclusion}.

\section{CMB Spectral Distortions}
\label{Sec: Distortions}

In this section, we describe how our interaction gives rise to CMB spectral distortions. The effect of our interaction, in the presence of dark matter, is to convert photons into a dark sector particle, $X$. For example, if the axion constitutes the dark matter, photons will interact with the axions and convert into dark photons so that $X$ is the dark photon. Conversely if the dark photon is dark matter, the photon will convert into axions and, in this case, $X$ is the axion. In order to speak generally about either axion or dark photon dark matter, we will refer to the particle the photon converts to as $X$ throughout this paper. The implications for the CMB spectral distortion are the same in either scenario since the important effect is that the photons convert to an invisible dark sector particle $X$ and are removed from the photon spectrum. 

The removal of any photons from the bath can lead to a deviation from the blackbody spectrum. We can quantify that change by a frequency dependent distortion $\delta(\omega)$ defined in Eq.~\ref{Eq: Distortion Definition}. 
\bea
\label{Eq: Distortion Definition}
f(\omega)=\frac{1}{e^{\omega/T}-1}\rar\frac{1}{e^{\omega/T}-1}\(1-\delta(\omega)\) \, .
\eea
The exact frequency dependence of the distortion will depend on when in cosmic history the photons were removed from the photon spectrum.  

\begin{figure}[h]
    \centering
    \includegraphics[width=0.95\textwidth]{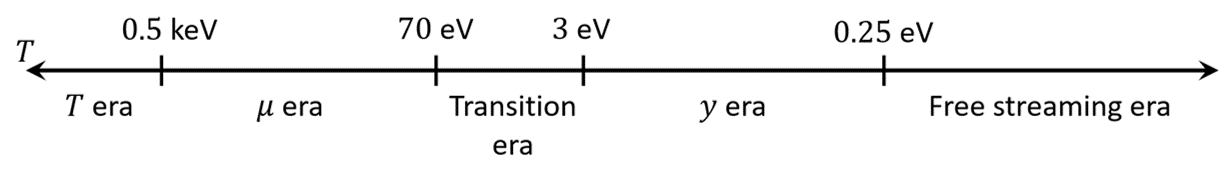}
    \caption{ A timeline of the types of relevant distortion eras. The timeline is presented with time described by decreasing temperature $T$.}
    \label{Fig: Distortion Summary}
\end{figure}

As shown in Figure~\ref{Fig: Distortion Summary}, we can define 5 different eras, the $T$ era, the $\mu$ era, the $\mu-y$ transition era, the $y$ era, and the free streaming era, in which the injection or removal of photons gives rise to different distortions. In the remaining of this section, we briefly review these different eras and discuss the characteristic effect of photon removal in each. As we will discuss, the final distortion to the blackbody spectrum can be parameterized by the impacts coming from different eras as
\bea
\label{Eq: Total Distortion}
\delta_{\text{Tot}}(\omega)=(\bar \mu +\bar \mu_t)M(\omega/T)+(y+y_t)Y(\omega/T)+\delta_{Doppler}(\omega)+\delta_{\text{Free}}(\omega).
\eea
We see there are 4 different types of distortions. The $\mu$ and $y$ distortions have distinct shapes and are insensitive to the details of the model generating the distortion, while $\delta_{Doppler}$ and $\delta_{free}$ have a model dependent shape. The contribution of all the pre-recombination distortions are computed using the Green's function method described in Ref.~\cite{Chluba:2015hma}, using the rate at which photons are converted into particle $X$, $\Gamma_{\gamma\rar X}(\omega)$, as discussed in Sec.~\ref{Sec: Computing Distortion} and Appx.~\ref{App: Green's Function}. In computing this rate, we will need to calculate is the conversion probability $P_{\gamma\rar X}(\omega)$. We also show that the distortion due to the post-recombination free streaming era distortion is directly related the the conversation probability $P_{\gamma\rar X}(\omega)$. Thus, the central quantity we will need to compute all spectral distortions is the conversion probability. 

\subsection{$T$ Era}

\begin{figure}[h]
    \centering
    \includegraphics[width=0.95\textwidth]{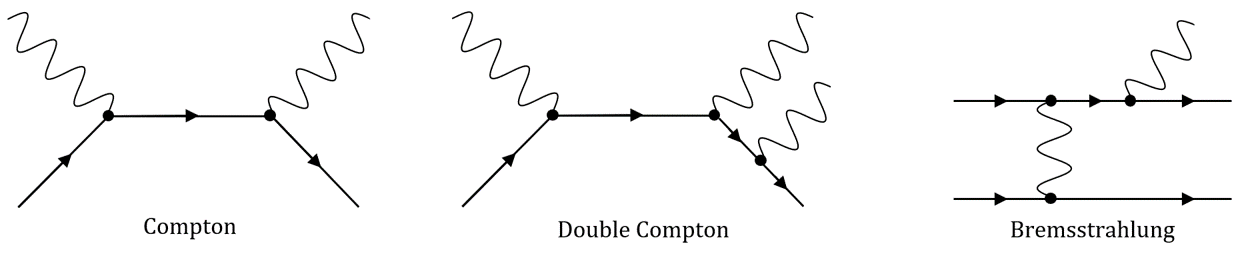}
    \caption{A sample diagram for each of the processes holding photons in equilibrium with the electrons.}
    \label{Fig: Photon-Electron Scattering}
\end{figure}

In the very early universe, at temperatures $T \gtrsim 0.5$ keV (redshifts $z\gtrsim 2\cdot 10^6$), a number of scattering processes involving photons are very efficient at driving the photon distribution towards an equilibrium distribution with zero chemical potential. The main processes, shown in Fig.~\ref{Fig: Photon-Electron Scattering}, are Compton scattering, which can quickly redistribute the photon energy and equilibrate the electron and photon temperatures, as well as number changing processes, such as double Compton and bremsstrahlung. Due to these processes, in this era, even if photons are lost due to conversion into X, the distribution would be quickly driven back to that of a blackbody and the only effect would be a small change in the blackbody temperature. Since we don't know a priori the temperature of the CMB, injections in this era would not lead to a bound from the CMB spectrum (there could be bounds by studying the anisotropies or comparing to Big Bang Nucleosynthesis predictions which we will not explore in this work).

\subsection{$\mu$ Era}
\label{Sec: mu dist}
Once the universe has cooled past $T\approx 0.5$ keV ($z\approx 2\cdot 10^6$), the higher order processes like double Compton scattering and bremmstrahlung are no longer efficient at setting the chemical potential to zero (although they can still be efficient for absorbing/emitting very low frequency photons). In this era, Compton scattering is still efficient at redistributing the energy, which drives the spectrum towards an equilibrium distribution. Because photon number is now conserved, any removal of energy will result in a small effective chemical potential term $\bar \mu$
\bea
\label{Eq: mu distortion}
f(\omega)=\frac{1}{e^{\omega/T}-1}\rar \frac{1}{e^{\omega/T+\bar \mu}-1}\approx \frac{1}{e^{\omega/T}-1}\(1-\bar \mu \, M(\omega/T)\).
\eea
This distortion has a fixed shape, $M(\omega/T)$, given in  Appx.~\ref{App: Green's Function}. The size of this distortion is captured by the effective chemical potential, $\bar \mu$, which can be calculated from $\Gamma_{\gamma\rar X}$, the rate at which photons are being converted to dark sector particles $X$ as shown in  Appx.~\ref{App: Green's Function}. COBE-FIRAS placed a bound of $|\bar \mu|<9\cdot 10^{-5}$~\cite{Fixsen:1996nj}.

\subsection{$y$ Era}
\label{Sec: y dist}
At temperatures lower than $3$ eV ($z\approx 10^4$), Compton scattering is still efficient enough to trap photons, but is now inefficient at changing photon energy, and transferring energy between photons and electrons. This leads to two effects. Firstly, some small amount of energy can be still exchanged with the electrons, leading to a difference in the photon and electron temperatures. Subsequent scatterings of photons with an electron fluid at a different temperature, lead to a $y$-distortion via the Sunyaev-Zeldovich (SZ) effect~\cite{Zeldovich:1969ff}. Secondly, energy injections/removals in a given frequency, can still be smeared due to Doppler broadening via Compton scattering, even if the process is no longer efficient at thermalizing the spectrum. This leads to two separate distortions: a $y$-distortion and a Doppler smeared distortion,
\bea
\label{Eq: y distortion}
f(\omega)=\frac{1}{e^{\omega/T}-1}\rar \frac{1}{e^{\omega/T}-1}\(1-y Y(\omega/T)-\delta_{Doppler}(\omega)\) \, .
\eea
The $y$-distortion has a fixed shape, $Y(\omega/T)$, given in  Appx.~\ref{App: Green's Function} and a size determined by the small parameter $y$ which can be computed from the photon loss rate $\Gamma_{\gamma\rar X}$ as described in  Appx.~\ref{App: Photon Rate}. COBE-FIRAS placed a constraint $|y|<1.5\cdot 10^{-5}$~\cite{Fixsen:1996nj}. On the other hand, the shape of the Doppler smeared distortion is model dependent, so we instead place a bound by comparing directly to the COBE-FIRAS data.

\subsection{$\mu-y$ Transition Era}
Once the temperature decreases below $T\approx 70$ eV ($z\approx 3\cdot 10^5$), Compton scattering, while still efficient at trapping photons, begins to become inefficient at redistributing energy for certain frequency modes of the photon spectrum. This signals the end of the $\mu$ era and the start of the $\mu-y$ transition era which lasts until $T\approx 3$ eV ($z\approx 10^4$). In this transition era, higher energy modes still redistribute energy efficiently through Compton scattering, while energy redistribution is inefficient for lower energy modes. At intermediate modes, energy redistribution is not efficient but is non-negligible. In order to exactly treat this very frequency dependent behavior, one would need to simulate the distortion numerically~\cite{Chluba:2015hma}. However, as noted in Ref.~\cite{Chluba:2015hma}, for the range of photon frequencies we are interested in, the distortion can be modeled to good accuracy as a pure energy injection as described in Ref.~\cite{Chluba:2013vsa}. The resulting spectral distortion is a combination of a $\mu$ distortion $M(x)$ and a $y$ distortion $Y(x)$ the shapes of which are given in  Appx.~\ref{App: Green's Function}.
\bea
\label{Eq: muy transition distortion}
f(\omega)=\frac{1}{e^{\omega/T}-1}\rar\frac{1}{e^{\omega/T}-1}\(1-\bar \mu_t \, M(\omega/T)-y_t \, Y(\omega/T)\).
\eea
The subscripts $t$ on the coefficients $\bar\mu_t$ and $y_t$ denote that these coefficients are calculated differently from  $\bar \mu$ in Eq.~\ref{Eq: mu distortion} and $y$ in Eq.~\ref{Eq: y distortion}. They still however are calculated from the photon conversion rate $\Gamma_{\gamma\rar X}$ as shown in Appx.~\ref{App: Green's Function}.

\subsection{Free Streaming Era}
\label{Sec: Free Streaming Era}
Around $T=0.25$ eV ($z=1100$), most electrons have been captured to form neutral hydrogen, and the universe becomes transparent to photons. From this point on, the photons become free streaming and can travel unimpeded across the universe, giving rise to the CMB we observe today. However, the presence of our interaction leads to a probability $P_{\gamma\rar X}(\omega)$ that a CMB photon with frequency $\omega$ will convert to an invisible dark sector particle $X$ before reaching us. Because the photons are free streaming, there is no thermalization, or redistribution of energy. So, the resulting spectrum is the original spectrum multiplied by the survival probability of a photon to reach us without converting to X, 
\bea
\label{Eq: free distortion}
f(\omega)=\frac{1}{e^{\omega/T}-1}\rar \frac{1}{e^{\omega/T}-1}\(1-P_{\gamma\rar X}(\omega)\) \, .
\eea
We can see that the distortion is simply the conversion probability $P_{\gamma\rar X}(\omega)$. The frequency dependence of this distortion is model dependent and as such we will have compute it and then constrain it directly with the COBE-FIRAS data to obtain a bound.

\section{Transition Probability}
\label{Sec: Probability}
\begin{figure}[h]
    \centering
    \includegraphics[width=0.95\textwidth]{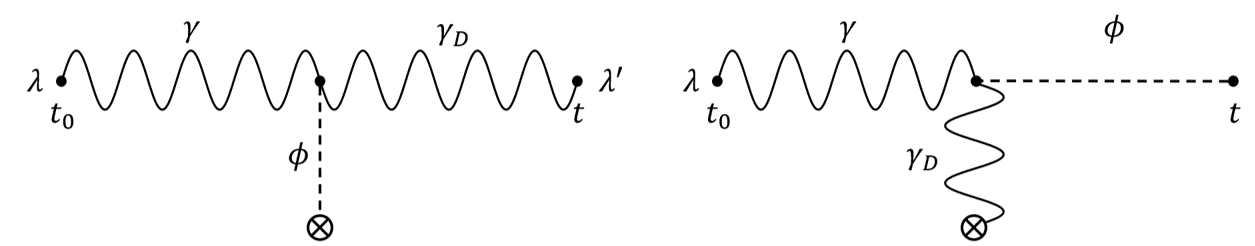}
    \caption{Diagrams for the probability of a photon produced at time $t_0$ to have converted to a dark photon or axion by a time $t$. The vertex indicates the interaction with the background dark matter field. $\lambda$ and $\lambda'$ represent the polarizations of the relevant particles. }
    \label{Fig: Photon Conversion}
\end{figure}

As described in the previous section, to compute the distortions we will need to compute the probability, $P(\omega,t,t_0)$, of converting photons of frequency $\omega$ produced at time $t_0$ into dark sector particles X at some later time $t$. Because our dark matter is made of bosons of mass $m\lesssim$ meV, the number density is large enough to treat dark matter as a classical background field. Therefore, these probabilities can be computed using Feynman diagrams like the one shown in Fig.~\ref{Fig: Photon Conversion}. 

In many instances the time interval $t-t_0$ will be of cosmic scale. For example, when applied to the free streaming distortion, $t-t_0$ will be the time between recombination and the present. As a consequence, we will need to compute these Feynman diagrams in a curved FRW background. We will work in conformal coordinates, described by the metric
\bea
\label{Eq: FRW metric}
d\tau^2=a^2(\eta)\(d\eta^2-d\vec x^2\) \, .
\eea
 In a slowly expanding universe ($m_{DM}, T_{CMB}\gg H$), we can easily expand any general scalar field $\phi$ and vector field $A_\mu$ in terms of ladder operators by solving their equations of motion using the WKB approximation. The details of this process are given in  Appx.~\ref{App: FRW QFT} and the result is

\bea 
\label{Eq: FRW Scalar Field}
\phi(x)=\int \frac{d^3\vec k}{(2\pi)^3a(\eta)\sqrt{2\omega^c(\eta)}}\(a_{\vec k}e^{-i\(\int ^\eta d\tilde \eta \omega^c(\eta,\vec k)-\vec k\cdot \vec x\)}+h.c.\)
 \eea
\bea
\label{Eq: FRW Vector Field}
A_\mu(x)=\int \frac{d^3\vec k}{(2\pi)^3\sqrt{2\omega^c(\eta)}}\sum_{\lambda=1,2,L}\(a^\lambda_{\vec k}\epsilon_\mu^\lambda(\vec k)e^{-i\(\int ^\eta d\tilde \eta \omega^c(\eta,\vec k)-\vec k\cdot \vec x\)}+h.c.\),
\eea 
where $a_{\vec k}$ ($a_{\vec k}^\lambda$) are the ladder operators of the scalar(vector) field, $\omega^c(\eta,\vec k)\equiv \sqrt{|\vec k|^2+a^2(\eta)m^2}$ is the conformal energy, and the polarizations $\epsilon_\mu^\lambda(\vec k)$ are given in Eq.~\ref{Eq: Polarizations}. 

We will also need to include plasma effects on the photon due to its impact on photon propagation and mixing. At all times of relevance, electrons are non-relativistic and lead to a plasma frequency, $\omega_p(\eta)$, given by
\bea
\label{Eq: Plasma Frequency} 
\omega_p^2(\eta)=\frac{4\pi\alpha_e}{m_e}n_e(\eta) \, ,
\eea
where $\alpha_e$ is the fine structure constant and $n_e(\eta)$ is the number density of electrons which changes as the universe expands. We are working in the limit where $\omega_p(\eta)$ is much smaller than the frequency of the CMB photons $\omega\sim T_{CMB}$ . In this limit, the effects of the plasma can be reduced to the photons acquiring a small mass $m_\gamma(\eta)=\omega_p(\eta)\ll T_{CMB}$. Despite the plasma frequency giving rise to an effective mass for transverse modes, at such large frequencies there are no longitudinal modes of the photon (plasmons)~\cite{Braaten:1993jw}. We use the redshift dependent plasma frequency from~\cite{Caputo:2020bdy}.

Due to the non-trivial time dependence of the fields given in Eq.~\ref{Eq: FRW Scalar Field} and Eq.~\ref{Eq: FRW Vector Field}, we will only Fourier transform the diagrams in space. In this regime, our diagrams are transition amplitudes with a time dependent interaction and so we can expand our amplitudes to leading order using a Dyson series, 
\bea 
\label{Eq: Transition Amplitude}
\bra X,\vec k'|U(\eta,\eta_0)|\gamma,\vec k'\ket \approx -i\int_{\eta_0}^\eta d\eta' \bra X,\vec k'|V_I(\eta')|\gamma,\vec k'\ket ,
\eea
$V_I(\eta')$ is our interaction potential given by 
\bea
\label{Eq: Interaction Pot}
V_I(\eta')=-\int d^3\vec x \, \LL_{int}=-\int d^3\vec x \,\frac{\phi}{2f_a}F_{\mu\nu}\tilde F^{\mu\nu}_D.
\eea
Note the momentum eigenstates in Eq.~\ref{Eq: Transition Amplitude} are normalized such that $\bra \vec k'|\vec k\ket =(2\pi)^3\delta(\vec k-\vec k')$ which differs from the usual Lorentz invariant normalization by a factor of $2\omega$. 

We can simplify $V_I(\eta')$ by using the fact that our dark matter field is nonrelativistic to ignore gradients in favor of time derivatives which simplifies the interaction to
\begin{equation}
\label{Eq: Interaction Pot Simplified}
\begin{aligned}
\text{Axion DM:} & \quad V^{\phi}_I(\eta')=\int d^3\vec x \,\frac{\partial_\eta \phi}{f_a}\vec A_D\cdot \vec B \\
\text{Dark Photon DM:} & \quad V^{D}_I(\eta')=-\int d^3\vec x \,\frac{\phi}{f_a}\partial_\eta \vec A_D\cdot \vec B.
\end{aligned}
\end{equation}
Next we insert the expansion of the fields for the photon's magnetic field $\vec B$ (Eq.~\ref{Eq: FRW Vector Field}) and the outgoing particle field $X$ (either Eq.~\ref{Eq: FRW Vector Field} for an outgoing dark photon or Eq.~\ref{Eq: FRW Scalar Field} for an outgoing axion) in terms of creation and annihilation operators. As discussed earlier, we will treat the dark matter as a classical background field. These potentials can be inserted in Eq.~\ref{Eq: Transition Amplitude} to compute the transition probability. After some simplifications described in  Appx.~\ref{App: FRW QFT}, this probability takes the form, 
\bea
\label{Eq: Prob Form}
P(\vec k,t_0,t)=\frac{1}{2}\sum_\lambda \bigg|\AA^\lambda (\vec k,t,t_0)\bigg|^2  \where \AA^\lambda(\vec k,t,t_0)=\int_{t_0}^t dt'\,\MM^\lambda(t',\vec k) \, ,
\eea 
with
\bea
\label{Eq: Prob Form A}
\MM^\lambda_{\gamma \rar \gamma_D}(t',\vec k)=\frac{\dot \phi(t',\tilde{\vec x}(t'))\sqrt{v_D(\vec k,t')}}{2f_a}
e^{ i\int^{t'}d\tilde t \(\omega_D(\tilde t)-\omega_\gamma(\tilde t)\)}
\eea 
\bea
\label{Eq: Prob Form DP}
\MM^\lambda_{\gamma \rar \phi}(t',\vec k)=\frac{\dot {\vec A}_D(t',\tilde{\vec x}(t'))\cdot \vec{\epsilon}_\lambda(\vec k)\sqrt{v_\phi(\vec k,t')}}{2a(t')f_a}
e^{ i\int^{t'}d\tilde t \(\omega_\phi(t)-\omega_\gamma(t)\)} \, ,
\eea 
where the first (second) equation corresponds to the axion (dark photon) dark matter scenario. In the above equations, $\tilde {\vec x}(t')$ is the position of the photon at time $t'$, the dots represent time derivatives with respect to comoving time $t$, and $v(\vec k,t)$ represents the velocity of a given particle. We have approximated $v_\gamma=1$. The $\omega$'s are now the physical energies defined as: 
\bea
\label{Eq: Physical Energies}
\omega(\vec k,t,m)=\sqrt{\frac{|\vec k|^2}{a^2(t)}+m^2}
\eea 
Finally, we must determine what form our dark matter background takes. By solving the equations of motion for the dark matter fields (Eq.~\ref{Eq: KG in FRW} and Eq.~\ref{Eq: Maxwell in FRW}) in the non-relativistic limit, keeping terms up to $\OO(v_{DM})$, and demanding that the energy density is $\rho_{DM}^0/a^3(t)$, where $\rho_{DM}^0$ is the energy density of dark matter of the universe at the present time, we find 
\bea
\label{Eq: Axion DM Field} 
\dot \phi_{DM}(\tilde{\vec x},t)=\sqrt{\frac{2\rho_{DM}^0}{a^3(t)}}\cos(m_{DM}t+\beta(\tilde{\vec x}))
\eea
\bea
\label{Eq: DP DM Field} 
\dot {\vec A}_{DM}(\tilde{\vec x},t)=\sqrt{\frac{2\rho_{DM}^0}{a(t)}}\vec \epsilon(\tilde{\vec x})\cos(m_{DM}t+\beta(\tilde {\vec x})).
\eea
Both fields get a spatially dependent phase, $\beta(\tilde{\vec x})$, while the vector dark matter field gets an additional spatially dependent polarization unit vector $\vec \epsilon(\tilde{\vec x})$. Both of these quantities vary spatially on the scale of the dark matters de Broglie wavelength $(m_{DM}v_{DM})^{-1}$ with $v_{DM}\ll 1$. Additionally, they vary in time, on timescales  $(m_{DM}v_{DM}^2)^{-1}$. Since the time dependence is suppressed by a factor $v_{DM}$ relative to the spatial dependence it will be ignored. In  Appx.~\ref{App: Avgs}, we show that for all distortions, we average $\tilde{\vec x}$ over many de Broglie wavelengths of the dark matter field, which means that we can average all of these spatially dependent quantities. We will leave the averaging over the phase for later in the computation, but in  Appx.~\ref{App: Avgs} we show that we can effectively replace 
\bea 
\label{Eq: VDM Polarization Average}
\vec \epsilon_\lambda(\vec k)\cdot \vec \epsilon(\tilde{\vec x})\rar \frac{1}{\sqrt{3}}.
\eea 

Physically, this factor $\frac{1}{\sqrt{3}}$ is reflecting the fact that the interaction $\vec\epsilon_\lambda(\vec k)\cdot \vec \dot{\vec A}_D$ in Eq.~\ref{Eq: Prob Form DP} is picking out one particular polarization of the vector dark matter. After averaging over $\tilde{\vec x}$, this particular polarization must make up $1/3$ of the total dark matter by isotropy, effectively sending $\rho_{DM}^0\rar \rho_{DM}^0/3$. 
Using Eqs.~\ref{Eq: Axion DM Field}-\ref{Eq: VDM Polarization Average} to simplify Eqs.~\ref{Eq: Prob Form}-\ref{Eq: Prob Form DP}, we can write the transition probabilities as 
\bea
\label{Eq: Probabilities}
&&P_{\gamma \rar \gamma_D}(\vec k,t_0,t)=\frac{\rho_{DM}^0}{2f_a^2}L^2(m_\phi,m_D,\vec k,t_0,t) \\ &&P_{\gamma \rar \phi}(\vec k,t_0,t)=\frac{\rho_{DM}^0}{6f_a^2}L^2(m_D,m_\phi,\vec k,t_0,t),\nn
\eea
where $L$ is a length scale defined as 
\bea
\label{Eq: L Def}
L^2(m_{DM},m_X,\vec k,t_0,t) \equiv && \bra \bigg|\int_{t_0}^t d\tilde t'\sqrt{\frac{v_X(t',|\vec k|)}{a^3(t')}}\cos(m_{DM}t'+\beta(t')) \right. \\ && \left. \times e^{i\int^{t'}d\tilde t \Delta \omega_{\gamma\rar X}(\tilde t,\vec k)}\bigg|^2 \ket_\beta \, ,\nn
\eea
where $\Delta \omega_{\gamma\rar X}$ is the change in energy from a photon converting into particle $X$ at momentum $\vec k$,
\bea
\label{Eq: Delta Omega}
\Delta \omega_{\gamma\rar X}(\tilde t,\vec k)=\omega_X(\tilde t,\vec k) -\omega_\gamma(\tilde t,\vec k) \, .
\eea
The $\bra\ket_\beta$ indicates the remaining average over the dark matter phase $\beta(t')=\beta(\tilde{\vec x}(t'))$ which is handled for pre-recombination and free streaming distortions separately in Appx.~\ref{App: Photon Rate} and~\ref{App: Free Dist Comp} respectively.
From Eq.~\ref{Eq: Probabilities}, we can see that the only difference between scalar dark matter and vector dark matter is the overall factor of $1/3$ in the conversion probability from the effect described above. This means that the coupling to dark photon matter is effectively $1/\sqrt{3}$ that of the coupling to axion dark matter and so the bounds placed on the coupling in the dark photon dark matter will be weaker than the bounds for the axion dark matter by a factor $\sqrt{3}$ . For simplicity, we will only consider axion dark matter going forward, knowing that we can translate any result to dark photon dark matter by multiplying by $1/\sqrt{3}$.

\section{Computing the Distortions}
\label{Sec: Computing Distortion}

In this section, we will use Eq.~\ref{Eq: Probabilities} to determine the strength of the distortions arising from the various eras. This will be very different for distortion generated pre-recombination versus in the free streaming regime, so we consider them separately. 
\subsection{Free Streaming Distortion}
\label{Sec: Free Streaming Dist}
In Sec.~\ref{Sec: Free Streaming Era}, we showed that the free streaming distortion, $\delta_{free}$, is equal to the probability of converting the photon to dark sector particle $X$ between recombination, $t_0$ and today, $t$. Thus, we need to compute
\bea
\label{Eq: Delta Free Gen}
\delta_{free}(|\vec k|)=\frac{\rho_{DM}^0}{2f_a^2}L^2(m_{DM},m_X,\vec k,t_0,t) \, .
\eea
From Eq.~\ref{Eq: L Def}, it is clear that we can write $L^2=\bra\big|\frac{L_++L_-}{2}\big|^2\ket_\beta$, where
\bea
\label{Eq: Lpm Def}
L_\pm=\int_{t_0}^t dt'\sqrt{\frac{v_X(t')}{a^3(t')}}e^{i\int^{t'}d\tilde t\, (\Delta\omega(\tilde t)\pm (m_{DM}-\dot\beta(\tilde t))}.
\eea
The remaining integral is an oscillatory integral with frequency 
\bea\label{Eq: Omegapm Def}
\Omega_\pm(t)\equiv \Delta \omega(t)\pm (m_{DM}-\dot\beta(t)).
\eea
Notice that all of the  time dependent quantities change on the Hubble scale $H(t)$ due to the expansion of the universe\footnote{Strictly speaking, $\dot\beta(t)$ changes on the scale $m_{DM}v_{DM}\delta v_{DM}\sim m_{DM}v_{DM}^2$ where $\delta v_{DM}$ is the size of the dark matter velocity dispersion. However, in the full computation one can work in Fourier space and treat each mode of the dark matter field independently before summing over all modes at the end, effectively removing effects from the time dependence from $\dot \beta(t)$.}

There are two limits in which this integral can be computed. The first is the fast oscillation limit where the oscillation frequency $\Omega_\pm(t)$ is approximately constant over many oscillations. This is the limit where 
\bea 
\label{Eq: Fast Cond}
\Omega_\pm(t)\gg H(t).
\eea
 In this limit, all of the time dependent quantities in Eq~\ref{Eq: Lpm Def} become approximately constant up to corrections of order $H/\Omega_\pm$ and the integral can be computed analytically. The second limit is the resonant limit. In this limit, there is a time  (or possibly multiple times), $t_r$, where there is a stationary phase in the exponential ($\Omega_\pm(t_r)=0$). Since $\dot \beta(t)\ll m_{DM}$ to a good approximation these resonant times can be found by solving the equation
\bea
\label{Eq: Resonant Condition}
\Delta\omega(t_r)\pm m_{DM}=0.
\eea
Physically, this corresponds to times in which the dark matter particle being absorbed/emitted by the photon is on shell, leading to an enhancement in the conversion probability. The stationary phase approximation is used to compute the integral in this limit. Appx.~\ref{App: Free Dist Comp} contains the details of computing the distortions in both of these limits. In the end, we find, 

\bea 
\label{Eq: Free Dist Final}
\delta_{free}(|\vec k|)=\frac{\rho_{DM}^0}{4f_a^2}\frac{|L_+|^2+|L_-|^2}{2}\where |L_\pm|^2= \begin{cases} \frac{v_X(a_*)}{a^3_*|\Delta\omega(a_*)\pm m_{DM}|^2}& \text{no resonances}\\
\sum_{a_{\pm}} \frac{2\pi v_X(a_\pm)}{a^3_\pm|\partial_t \Delta\omega(a_\pm)|}& \text{resonances}.
\end{cases}
\eea 
Here, $a_\pm$ are all solutions to Eq.~\ref{Eq: Resonant Condition}, and $a_*=(1090)^{-1}$ is the redshift at recombination. Notice that the resonant distortion is enhanced by a factor of $\frac{|\Delta\omega\pm m_{DM}|^2}{\partial_t \Delta\omega}\sim \frac{m_{DM}}{H}$ with respect to the non-resonant distortion. Even for the smallest possible dark matter masses, $m_{DM}\sim 10^{-20}$, this is an enhancement by a factor of $\OO(10^8)$. Thus, we expect our bounds on the coupling $1/f_a$ to be enhanced by orders of magnitude in regions of parameter space where these resonances happen. 

\subsection{Pre-Recombination Distortions}

The application of Eq.~\ref{Eq: Probabilities} to pre-recombination distortions is not as straightforward as for the free streaming distortions. The distortions can be computed via the Green's function method outlined in Ref.~\cite{Chluba:2015hma} where the distortion, $\delta(x)$ is given by
\bea 
\label{Eq: Green's function}
\delta(x)=\int dx' \int \frac{da}{aH(a)} G(x,x',a)\Gamma_{\gamma \rar X}(x',a),
\eea 
where $x$ is the dimensionless frequency $x=\omega/T(a)$ and $\Gamma_{\gamma \rar X}(x',a)$ is the rate at which photons of frequency $x'T$ are converted to dark sector particle $X$. The Green's function $G(x,x',a)$ describes how photons injected into mode $x'$ at time when the scale factor is $a$ are redistributed to mode $x$. $G(x,x',a)$ is given for the various eras in  Appx.~\ref{App: Green's Function}. In this section we will describe how to compute the rate $\Gamma_{\gamma \rar X}(x',a)$ appearing in Eq.~\ref{Eq: Green's function}. 

We can compute $\Gamma(a,\vec k)$ from $P(\vec k,t_0,t)$ as follows. Consider a photon that scatters off of an electron at time $t_0$ and travels some time $\tau$ before scattering off of another electron. $P(\vec k,t_0,t_0+\tau)$ is then the probability of the photon converting to $X$ between these scatterings. For an ensamble of photons scattering with time $\tau$ between scatterings, the rate at which those photons are converted to $X$ is then 
\bea 
\label{Eq: Unavg rate}
\Gamma(t_0,\tau, \vec k)=\frac{P(\vec k,t_0,t_0+\tau)}{\tau}.
\eea
Since the photon is traveling through a very dense medium of electrons, it has a certain probability $p(\tau)$ of traveling a distance $\tau$ characterized by its mean free path $\lambda_\gamma$, given by
\bea
\label{Eq: mfp prob}
p(\tau)=\frac{e^{-\tau/\lambda_\gamma}}{\lambda_\gamma}.
\eea
In order to find the average rate for all photons, we average $\Gamma(t_0,\tau, \vec k)$ over this path length distribution
\bea
\label{Eq: Rate averaged}
\Gamma(a(t_0),\vec k)=\int_0^\infty d\tau \,\frac{e^{-\tau/\lambda_\gamma}}{\lambda_\gamma}\frac{P(\vec k,t_0,t_0+\tau)}{\tau}.
\eea
Because this era is before recombination, we have $H\lambda_\gamma\ll 1$. In this limit, we can treat space as static and all parameters that change due to the expansion of the universe as constant in the integral, and Eq.~\ref{Eq: Rate averaged} can be computed analytically. The details are given in  Appx.~\ref{App: Photon Rate}, and the end result is
\bea
\label{Eq: Rate final}
\Gamma(a,|\vec k|)=\frac{\rho_{DM}^0v_X(a,|\vec k|)}{4f_a^2a^3\lambda_\gamma(a)}L_{eff}^2(a,|\vec k|), \eea
where
\bea
\label{Eq: Leff}
L_{eff}^2(a,|\vec k|) & = & \frac{1}{2}\left\{  \frac{\ln [1+\lambda_\gamma^2(a)(m_{DM}+\Delta\omega(a))^2]}{(m_{DM}+\Delta\omega(a))^2} \right. \\   & & + \left. \frac{\ln [1+\lambda_\gamma^2(a)(m_{DM}-\Delta\omega(a))^2]}{(m_{DM}-\Delta\omega(a))^2} \right\}.\nn
\eea
This rate can be plugged into Eq. ~\ref{Eq: Green's function} and integrated numerically using the Green's functions given in App~\ref{App: Green's Function} find the distortion from the various pre-recombination eras.

\section{Results}
\label{Sec: Constraints}
\begin{figure}[h!]
    \centering
\includegraphics[width=.49\textwidth]{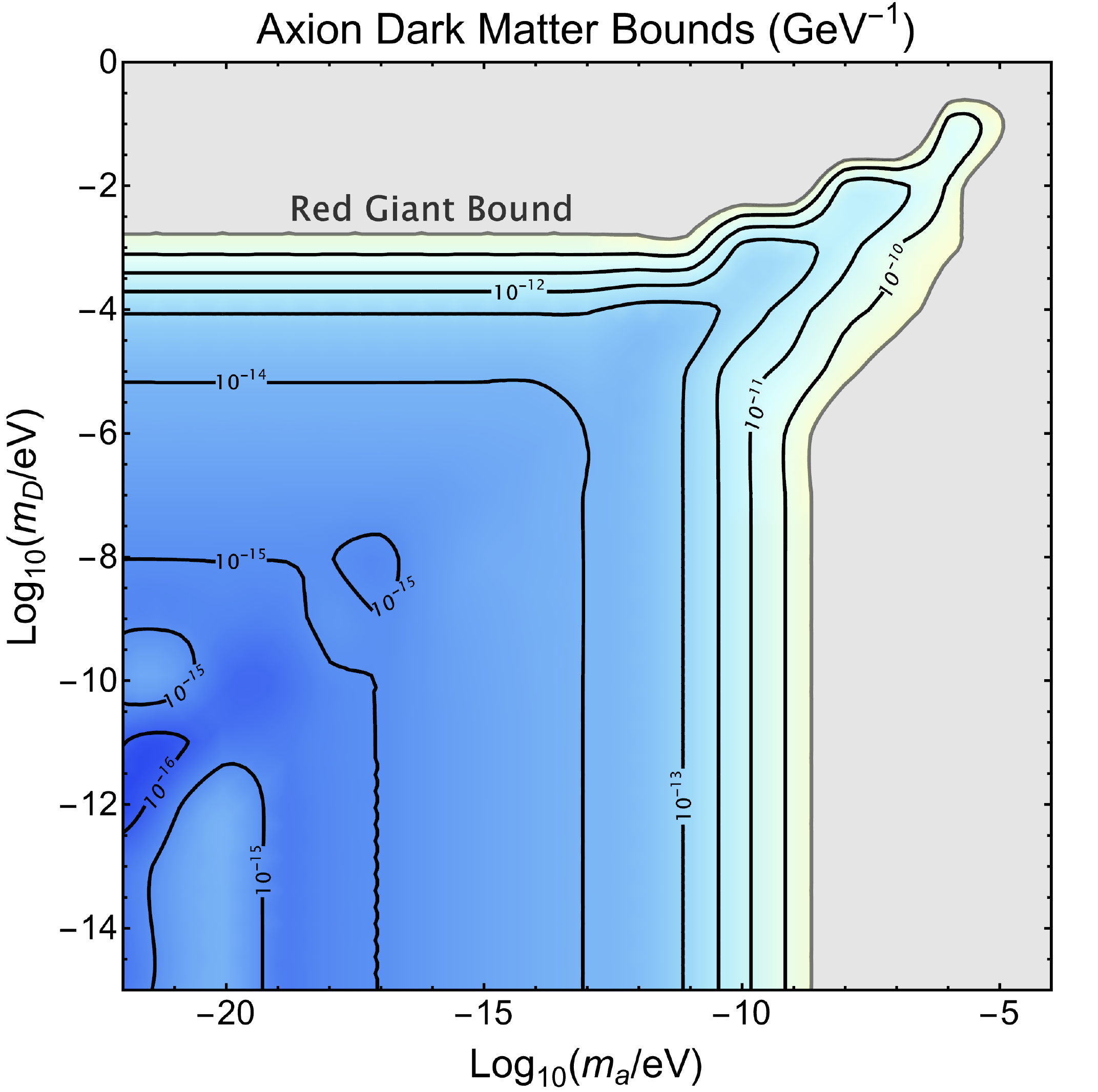}
\includegraphics[width=.49\textwidth]{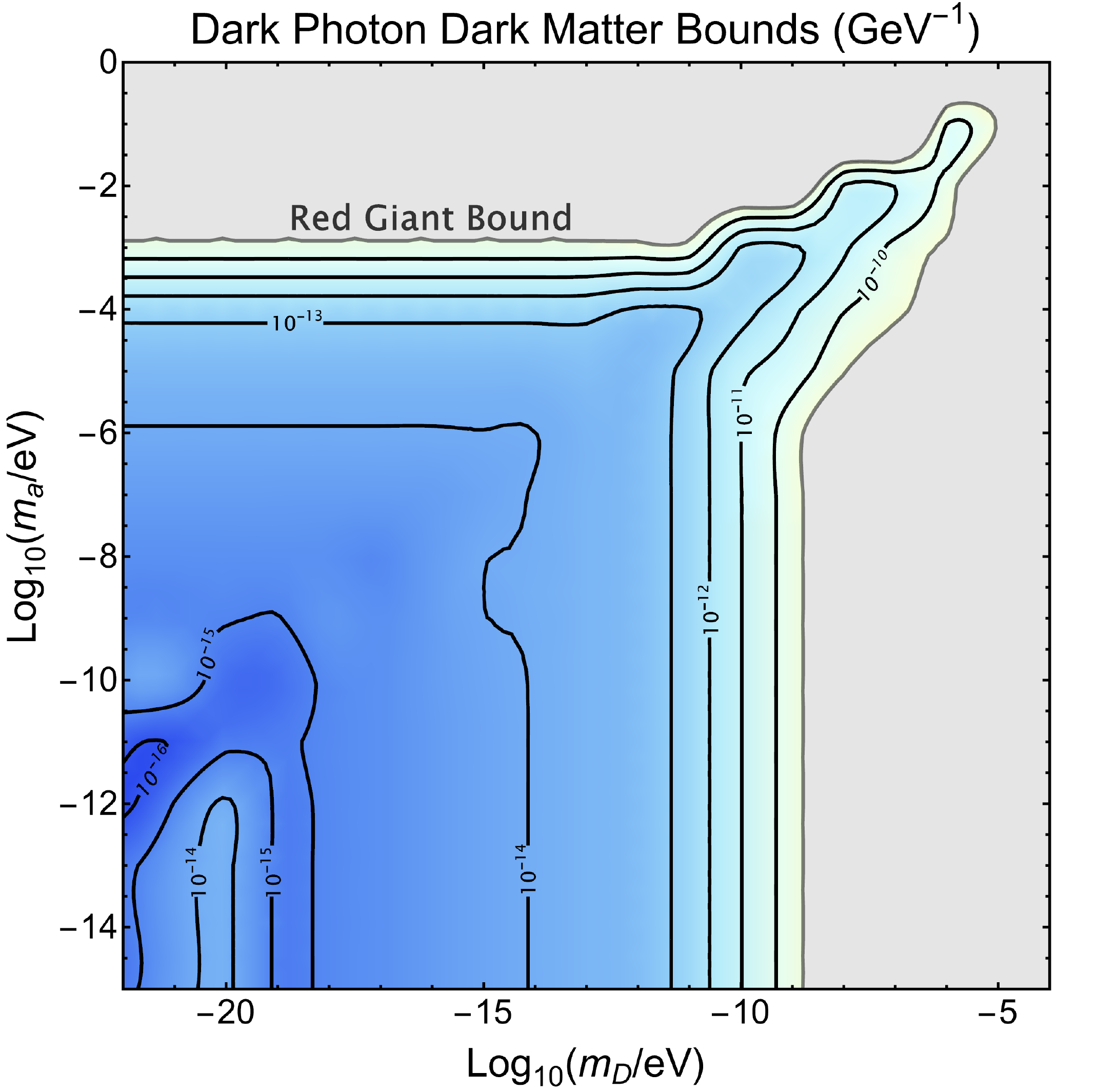}
    \caption{Bounds on the coupling $1/f_a$ (GeV$^{-1}$) plotted as a contour plot as a function of the dark photon mass ($m_D$) and the axion dark matter mass ($m_a$). The plot on the left shows the bounds for the axion dark matter case and the plot of the right shows the dark photon dark matter case. The grey region represents the region where the previous best bound on the coupling derived from red giant cooling~\cite{Carenza:2023qxh} is stronger.}
    \label{Fig: Axion Contour Bounds}
\end{figure}

\begin{figure}[h!]
    \centering
    \includegraphics[width=1\textwidth]{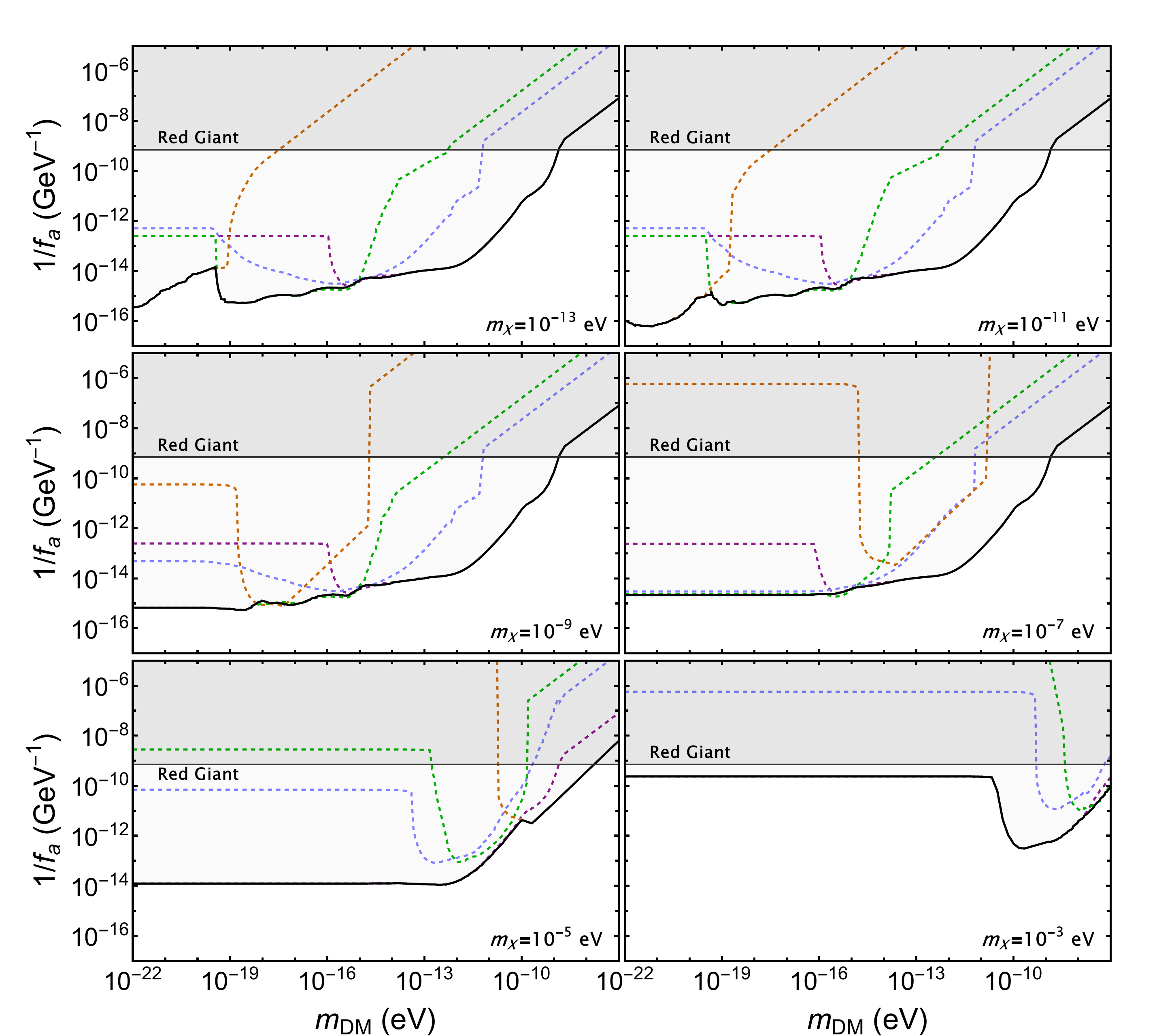}
    \caption{Bounds on the coupling $1/f_a$ as a function of the axion dark matter mass $m_{DM}$ plotted for various values of the dark photon mass, $m_X$. The purple, blue, green and orange dashed lines are the individual $\mu$, $y$, Doppler and free streaming bounds respectively while the solid black line is the total bound from all distortions. The grey line represents the current best bound on the coupling derived from red giant cooling~\cite{Carenza:2023qxh}. The bounds shown here are for the axion dark matter case, but the bounds for the dark photon dark matter case can be obtained by scaling the bounds up by a factor of $\sqrt{3}$. The bounds for  $m_X<10^{-13}$ eV are exactly those given in the upper-left plot.}
    \label{Fig: Bounds}
\end{figure}

The total distortion for a given set of masses $m_X$ and $m_{DM}$ and coupling $1/f_a$ is given by Eq.~\ref{Eq: Total Distortion}. As described in Sec.~\ref{Sec: Distortions}, temperature shift distortions are undetectable by COBE-FIRAS. Therefore we should add an arbitrary temperature shift, $\TT(x)$ (defined in App.~\ref{App: Green's Function}), to the distortion and do a best fit to COBE-FIRAS data~\cite{Fixsen:1996nj} with both the coupling, $1/f_a$ and the size of the temperature shift, $\alpha$, as free parameters. 
\bea
\label{Eq: Total Distortion w/ subtract}
\delta_{\text{Tot}}(\omega)=(\bar \mu +\bar \mu_t)M(\omega/T)+(y+y_t)Y(\omega/T)+\delta_{Doppler}(\omega)+\delta_{\text{Free}}(\omega)-\alpha \TT(\omega/T).
\eea
 However we can simplify this by demanding that the number density of CMB photons be the the same as that of a perfect blackbody at the measured temperature $T_{CMB}=2.35 \cdot 10^{-4}$ eV. This is exactly the procedure that is commonly done for $\mu$ and $y$ distortions as described in~\cite{Lucca:2019rxf}. This constraint fixes the size of the temperature distortion $\alpha$. For fixed $m_D$ and $m_\phi$, we do a $\chi^2$ fit of our distortion to the COBE-FIRAS data with a single free parameter, $1/f_a$. By demanding that the distorted spectrum matches the measured spectrum to within $2\sigma$, we obtain bounds for the coupling as a function of the dark photon and axion masses $m_D$ and $m_\phi$. Fig.~\ref{Fig: Axion Contour Bounds} shows a contour plot of these bounds as a function of the axion and dark photon masses. We show contours for both axion dark matter and dark photon dark matter. These bounds are plotted against the leading best bound on this coupling from red giant cooling constraints. Ref.~\cite{Carenza:2023qxh} shows that this coupling leads to a novel cooling mechanism in red giants due to plasmon decay and deduced that the cooling from this coupling is equivalent to the cooling from a neutrino magnetic dipole moment $\mu_\nu=\frac{1}{2f_a}$. Then using the bound placed on neutrino magnetic dipole moments found in~\cite{PhysRevD.102.083007} from red giant cooling they were able to place a bound $1/f_a<7.1 
 \cdot 10^{-10}$ GeV on the axion-photon-dark photon coupling. As seen in Fig.~\ref{Fig: Bounds}, in a large portion of parameter space, roughly $m_X<10^{-2} $ eV  and $m_{DM}<10^{-8}$ eV, our bounds beat this red giant
 bound by several orders of magnitude. It is worth noting that while the red giant bound is the most stringent bound in this region of parameter space (aside from the bounds placed in this work), there have been numerous other constraints placed on the coupling in this region of parameter space. The bounds placed on this coupling from stellar evolution~\cite{Choi:2018mvk}, Horizontal Branch stars~\cite{Arias:2020tzl}, and white dwarfs~\cite{Hook:2021ous} are all with in an order of magnitude of those from red giants. For simplicity, we only include the red giant bound in our plots.

 Fig.~\ref{Fig: Bounds} shows our bounds as a function of the axion dark matter mass ($m_{DM}$) for selected values of the dark photon mass ($m_X$) and shows the contribution to these bounds from each distortion era. While the bounds in Fig.~\ref{Fig: Bounds} are shown for axion dark matter, the equivalent bounds for dark photon dark matter can be found by scaling the bounds up by a factor of $\sqrt{3}$ as discussed at the end of section ~\ref{Sec: Probability}. As can be seen from the colored dashed lines in Fig.~\ref{Fig: Bounds} each constraint from each distortion has roughly the same behavior: constant for small $m_{DM}$ and increasing linearly in $m_{DM}$ for large $m_{DM}$ with an enhanced region in between. We can understand why the bounds have this behavior. Firstly, the enhanced region is the region of parameter space where photons during that particular era are able to resonantly convert. The other two limits can be understood by considering the integral $L^2$ defined in Eq.~\ref{Eq: L Def} in the non-resonant regime. Here $L^2$, the effective oscillation length, is the square of an oscillatory integral and thus should scale as $L^2\sim \Omega^{-2}$ where $\Omega$ is the fastest oscillation frequency in the integral. For sufficiently small $m_{DM}$, the coherent oscillation of dark matter is unimportant.  As such, the oscillation length is the standard $\Delta\omega_{\gamma\rar X}$ present for well known systems such as neutrino oscillations.  For larger $m_{DM}$, the fast oscillation of dark matter dominates and the oscillation frequency is $m_{DM}$. This combined with Eq.~\ref{Eq: Probabilities}, shows that the conversion probability $P_{\gamma\rar X}\propto \frac{\rho_{DM}}{f_a^2\Delta\omega^2}$ for small $m_{DM}$ and $P_{\gamma\rar X}\propto \frac{\rho_{DM}}{f_a^2m_{DM}^2}$ for large $m_{DM}$. Since the distortions all scale with the conversion probability, its clear that the bounds at low $m_{DM}$ are independent of $m_{DM}$ and linearly proportional to $m_{DM}$ for large $m_{DM}$.

Given that the COBE-FIRAS data was collected over 30 years ago, current technology could measure the CMB spectrum to higher precision. In fact, there are proposals for experiments, like PIXIE~\cite{Kogut_2011}, that aim to measure the spectrum to within a factor of $10^{-8}-10^{-9}$, an improvement of around a 3 to 4 orders of magnitude from COBE-FIRAS. These future experiments could potentially measure a distortion in the CMB frequency spectrum and thus it is interesting to ask what such a measured distortion could tell us about our dark matter models. 

\begin{figure}[h]
    \centering
    \includegraphics[width=1\textwidth]{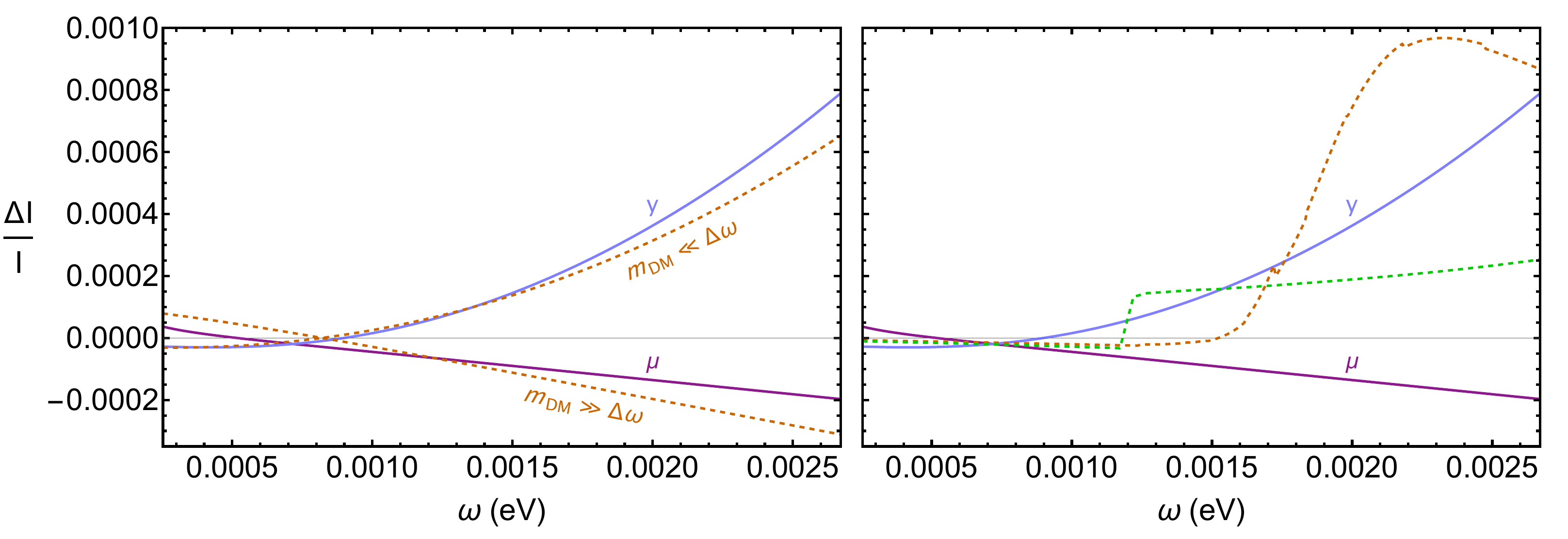}
    \caption{Shown here are shapes of the various types distortions $\delta(\omega) =\frac{I(\omega)-I_{Blackbody}(\omega)}{I_{Blackbody}(\omega)}$. Each distortion is plotted with an amplitude such that it disagrees with the COBE-FIRAS data at 2$\sigma$. Thus all the distortions are effectively the same strength. The purple and blue lines are $\mu$ and $y$ distortions respectively while the dashed green and orange lines depict Doppler and free-streaming distortion shapes for different choices of masses $m_{DM}$ and $m_X$. On the left, we show the distortions in the non-resonant limits given in Eq.~\ref{Eq: Free Dist Shape Non-res}. On the right, we show two of the many possible shapes the distortions can take when there is a resonance in either the Doppler or free distortions.   These resonant distortions have shapes distinct from the $\mu$ and $y$ distortions. It is also worth noting that the differences in shape between the Doppler and free distortion in the right-hand plot are due to the difference in choice of parameters, $m_{DM}$ and $m_X$, rather than a difference in distortion type. }
    \label{Fig: Distortion Shapes}
\end{figure}

Specifically, we will discuss qualitatively whether a distortion produced from our dark matter model(s) could potentially be distinguished from other distortion sources. Energy injection or removal into or from the background electron plasma before recombination produces primarily a $\mu$ and/or $y$ distortion and thus these types of distortions which are essentially model independent. However, because of their non-thermal origin, the Doppler distortion arising pre-recombination, and free streaming distortion post-recombination have model dependent spectral shapes and do provide a distinctive signature. For simplicity, we can focus on the free streaming distortion to get a sense of the various types of shapes this distortion can take. To start, one can take the large and small $m_{DM}$ limits ($m_{DM}\gg \Delta \omega$ and $m_{DM}\ll \Delta \omega$ respectively) of Eq.~\ref{Eq: Free Dist Final} and see that 
\bea
\label{Eq: Free Dist Shape Non-res}
\text{Small $m_{DM}$ limit:}\quad \delta\sim \omega^2 \qquad  \text{Large $m_{DM}$ limit:}\quad \delta\sim \text{Constant}.
\eea
Figure~\ref{Fig: Distortion Shapes} shows these quadratic and constant distortions plotted against the $\mu$ and $y$ distortions. The amplitudes of these distortions in Fig.~\ref{Fig: Distortion Shapes} are chosen so that each distortion disagrees with COBE-FIRAS at $2\sigma$. In this sense we can think of these distortions as being of equal strength. We can see that the large $m_{DM}$ matches very well with the $\mu$-distortion and the small $m_{DM}$ distortion, matches well with a  $y$ distortion making them difficult to distinguish from the generic $\mu$ and $y$ distortions respectively.

The shapes of the free streaming distortion get more interesting if we consider the resonant region of parameter space ($m_{DM}\sim \Delta \omega$). Here the distortion depends on the time of the resonance $a_\pm$ which is found by solving Eq.~\ref{Eq: Resonant Condition}. Because $\Delta\omega$ depends of the frequency of the photon, so does the resonant time through Eq.~\ref{Eq: Resonant Condition}. Thus different frequency modes can have different resonance times which can lead to very distinctive frequency dependencies in the distortion easily distinguishable from the standard $\mu/y$-distortion. In particular, it is possible that some frequency modes undergo resonance, while other modes don't. This leads to especially unique distortions, where some frequency modes are distorted while others, effectively, are not. Such extreme distortions are shown in the orange and green dashed lines in figure Fig.~\ref{Fig: Distortion Shapes}. As can be seen, lower frequency photons never resonate, and thus are effectively undistorted while higher frequencies do experience a distortion due to resonance. Additionally, the frequency dependence of the resonant piece of the distortion is not a simple power law of $\omega$ due to the dependence of the resonant time on the frequency and in turn the non-trivial dependence of the distortion on the resonant time. It is also important to note that these these qualitative features can also arise from the Doppler distortion. From Eq.~\ref{Eq: Rate final} and~\ref{Eq: Leff} one can derive the same small and large $m_{DM}$ behavior and show similar types of resonant behavior are possible. The green dashed line in Fig.~\ref{Fig: Distortion Shapes} shows one such distinctive resonant shape for the Doppler distortion. This makes the prospect of observing these unique distortions even more likely since, as shown in Fig.~\ref{Fig: Bounds}, there are significant regions of parameter space where the Doppler distortion leads to the strongest bound, which shows that in such regions of parameter space it is the most observable effect. It is worth pointing out that difference in shape and severity of the jumps of the Doppler and free distortions in Fig.~\ref{Fig: Distortion Shapes} is not due to an inherent difference between the free and Doppler distortion, but rather a difference in the parameters $m_{DM}$ and $m_X$ at which these distortions are evaluated. These set of parameters were chosen to highlight the difference in distortion shapes achievable by either the free or Doppler distortions rather than an inherent difference between them.

\section{Conclusion}
\label{Sec: Conclusion}

In this paper, we studied the effects from an axion-photon-dark-photon coupling to the Cosmic Microwave Background if either the axion or the dark photon is dark matter. This interaction, in a dark matter background, induces mixing of the photon with a new light boson, and can remove photons from the universe either before or after recombination. Removing photons from the baryon-photon plasma before recombination produces the well known, model independent, $\mu$ distortion or $y$ distortion, as well as a model dependent distortion due to Doppler broadening. The size of the distortion is determined by the rate at which photons are removed from the spectrum. Removing photons after recombination naturally changes the frequency spectrum of the CMB and thus produces a distortion with a new spectral shape. We computed these individual distortions and the corresponding total distortion that would be produced by an interaction of this type in the presence of either axion or dark photon dark matter. 

The distortion produced in the presence of dark photon dark matter is smaller than that produced by axion dark matter by a factor of $1/3$, but otherwise identical in terms of dependence on the model parameters. This is due to the energy of dark matter being spread over the 3 polarizations of dark photon dark matter, as opposed to the single polarization for axion dark matter. The interaction we study only couples a photon of given polarization to a single polarization of dark photon dark matter. Thus, a given photon effectively only couples to $1/3$ of the total dark matter background. This effectively leads to an interaction strength for dark photon dark matter that is $1/3$ of that for axion dark matter.

By comparing our computed distortions with the COBE-FIRAS data measuring the CMB frequency spectrum, we were able to place very restrictive bounds on our coupling $1/f_a$. These bounds are a significant improvement of several orders of magnitude over the previous best bound on this coupling from red giant cooling constraints~\cite{Carenza:2023qxh} as shown in Fig.~\ref{Fig: Bounds}. Additionally one can compare these bounds to those placed on the $\phi F\tilde F$ coupling in~\cite{Fedderke:2019ajk} and see that our bounds are several orders of magnitude stronger. 

We also briefly considered the possibility for future measurements of the CMB frequency spectrum to detect distortions produced by this model. While the $\mu$ and $y$ distortions are produced by any mechanism that adds or removes energy from the photon spectrum before recombination, the distortions generated in the free streaming era when the resonance condition is met lead to much more distinctive spectral features. Thus, such resonant distortions offer a promising avenue to single out the axion-photon-dark-photon interaction interaction if the next generation experiments measures distortions on the CMB spectrum.

\section*{Acknowledgements}
We would like to thank Asimina Arvanitaki, Junwu Huang and Ken Van Tilburg for very helpful and insightful discussion. AH, GMT, and CR are supported in part by the NSF under Grant No. PHY-2210361 and by the Maryland Center for Fundamental Physics (MCFP). GMT is also supported in part by the US-Israeli BSF Grant 2018236. This research was supported in part by Perimeter Institute for Theoretical Physics. Research at Perimeter Institute is supported by the Government of Canada through the Department of Innovation, Science and Economic Development and by the Province of Ontario through the Ministry of Research and Innovation.

\pagebreak 
 \appendix

 \section{A Model}
\label{App: Model}

Here we summarize a simple model containing an axion, photon and dark photon where the leading interaction is that given in Eq.~\ref{Eq: Interaction}. This model contains a dark sector with a complex scalar $\Phi$
a dark sector $U(1)$ gauge boson $A_D$ and two sets of two left handed Weyl fermions $\xi$, $\xi^c$, $\chi$ and $\chi^c$.
These fermions are charged under both electromagnetism and the dark $U(1)$ gauge group with charges shown in Table~\ref{Tab: Charges}. The scalar $\Phi$ is uncharged under both electromagnetism and the dark $U(1)$.

\begin{table}[h]
\centering
  \begin{tabular}{ | l || c | r |}
    \hline
     & $Q_{EM}$ & $Q_D$ \\ \hline \hline
    $\chi$ & 2 & 2 \\ \hline
    $\chi^c$ & -2 & -2 \\ \hline 
    $\xi$ & 2 & -2\\ \hline
    $\xi^c$ & -2 & 2\\ \hline
    \hline
  \end{tabular}
\caption{The charges for our dark sector fermions. $Q_{EM}$ is the particles electric charge in units of the fundamental electric charge and similarly, $Q_D$ is the particles charge in units of the fundamental dark charge.}
\label{Tab: Charges}
\end{table}
In addition to these charges we demand that our model obey a dark charge conjugation symmetry $C_D$ defined by 
\bea
\label{Eq: CD Sym}
C_D: \quad \quad A \rar A \qquad A_D\rar -A_D \qquad \chi \longleftrightarrow \xi\qquad \chi^c \longleftrightarrow \xi^c \qquad \Phi \rar\Phi^*
\eea 
and a $\mathbb{Z}_8$ symmetry defined by 
\bea
\label{Eq: CD Sym}
\mathbb{Z}_8: \quad \quad  \chi \rar e^{-i\pi/4}\chi \qquad\xi\rar e^{i\pi/4}\xi \qquad \Phi \rar e^{i\pi/4}\Phi.
\eea
This $\mathbb{Z}_8$, which is a subgroup of the would be $U(1)_{PQ}$ associated with the axion, may seem troublesome since it corresponds to a chiral rotation of the $\chi$ and $\xi$ by $-\pi/4$ and $\pi/4$ respectively and thus should generate an anomalous term through the chiral anomaly. However, with the charges defined in Table~\ref{Tab: Charges}, one can show that this is equivalent to a rotation of a single Weyl fermion with charges $Q_{EM}=Q_D=1$ by an angle of $2\pi$. Our theory should be consistent with the addition of a Dirac fermion with fundamental electric and dark charge and so a chiral rotation of this fermion by $2\pi$ must leave the theory invariant. Thus the $\mathbb{Z}_8$ symmetry is nonanomalous. Additionally, this $\mathbb{Z}_8$ symmetry forbids any operators in the potential $V(\Phi)$ that could give the axion a mass up to dimension 8. 

With these symmetries in mind, we write all possible terms in our Lagrangian up to dimension 4,
\bea
\label{Eq: Model Lagangian}
\LL=\LL_{SM}+\LL_{kinetic}-V(\Phi)+y(\Phi\chi\chi^c+\Phi^*\xi\xi^c)+h.c.\;,
\eea
where $V$ is the scalar potential, the last term is a Yukawa coupling for the fermions to the scalar and $\LL_{kinetic}$ contains all of the kinetic terms for the fermions, $\Phi$, and $A_D$ with the gauge couplings packaged in covariant derivatives. 

Now let us suppose that $\Phi$ under goes symmetry breaking and obtains a VEV,
\bea
\label{Eq: Phi vev}
\Phi(x)=(v+h(x))e^{i\phi(x)/f}.
\eea
Amongst other changes, the Yukawa piece becomes
\bea
\label{Eq: Yukawa Change}
y(\Phi\chi\chi^c+\Phi^*\xi\xi^c) \rar y(v+h)(e^{i\phi/f}\chi\chi^c+e^{-i\phi/f}\xi\xi^c).
\eea 
These phases can be eliminated by a chiral rotation,
\bea 
\label{Eq: Chiral Rotation}
\chi\rar \chi e^{-i\phi/f} \quad \quad \xi\rar \xi e^{i\phi/f}
\eea
which  naturally generates anomalous terms in the Lagrangian due to the chiral anomaly. It is a simple exercise to show that these terms are

\bea\label{Eq: Anomalous Terms}
\LL\supset \frac{\phi(x)}{16\pi^2f}\bigg(&\((eQ^\chi_{EM})^2 F_{\mu\nu}\tilde F^{\mu\nu}+2ee_DQ^\chi_{EM}Q^\chi_D F_{\mu\nu}\tilde F^{\mu\nu}_D+(e_DQ^\chi_{D})^2 F^D_{\mu\nu}\tilde F_D^{\mu\nu}\)  \\
& -\((eQ^\xi_{EM})^2 F_{\mu\nu}\tilde F^{\mu\nu}+2ee_DQ^\xi_{EM}Q^\xi_D F_{\mu\nu}\tilde F^{\mu\nu}_D+(e_DQ^\xi_{D})^2 F^D_{\mu\nu}\tilde F_D^{\mu\nu}\)\bigg).\nn
\eea
We get two terms, one for each rotation. Plugging in the charges given in Table~\ref{Tab: Charges} we can easily see that the $F \tilde F$ and $F_D\tilde F_D$ terms vanish while the $F\tilde F_D$ term remains
\bea 
\label{Eq: Our Coupling}
\LL\supset \frac{ee_D}{2\pi^2f} \phi F_{\mu\nu}\tilde F^{\mu\nu}_D.
\eea

\begin{figure}[h]
    \centering
    \includegraphics[width=0.95\textwidth]{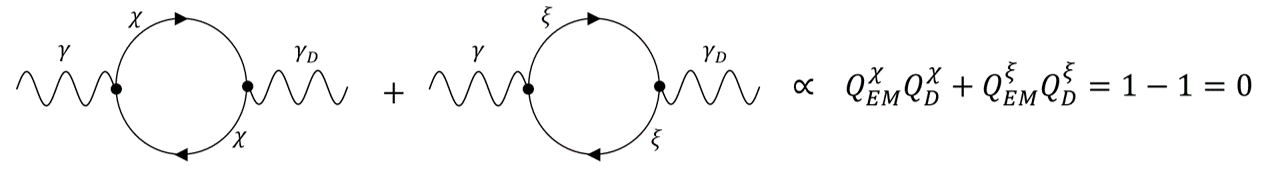}
    \caption{At 1-loop order, it is easy to see that the diagrams for kinetic mixing exactly cancel due to the $\chi$ and $\xi$ particles having opposite charges.}
    \label{Fig: Kinetic Mixing}
\end{figure}

A similar cancellation happens in the kinetic mixing term. At one loop, as shown in Fig.~\ref{Fig: Kinetic Mixing}, there are two diagrams for the kinetic mixing term which exactly cancel due to the $\chi$ and $\xi$ particle's opposite dark charges but identical masses from the $\Phi$ VEV. At the heart of this cancellation is the dark charge conjugation symmetry, $C_D$. This symmetry can easily be seen to forbid the generation of a kinetic mixing terms, which shows the cancellation observed at 1-loop occurs to all orders in perturbation theory.

\section{Computation of the conversion probability}
In this appendix we summarize the details of our conversion probability computation. 
\subsection{Quantized FRW Fields}
\label{App: FRW QFT}
Here we give a brief description of general massive scalar and vector fields in an expanding FRW background metric $d\tau^2=a^2(\eta)(d\eta^2-d\vec x^2)$. We can expand the fields in ladder operators with mode functions $u(\eta,\vec k)$ for the scalar and $v_\lambda^\mu(\eta,\vec k)$ for the vector where $\lambda$ is a polarization index.
\bea 
\label{Eq: FRW Ansatz}
&&\phi(x)=\int \frac{d^3\vec k}{(2\pi)^3} u(\eta,\vec k)a_{\vec k}e^{i\vec k\cdot \vec x}+h.c. \\
&& A_\mu(x)=\int \frac{d^3\vec k}{(2\pi)^3}\sum_\lambda\( v^\lambda_\mu(\eta,\vec k) a^\lambda_{\vec k}e^{i\vec k\cdot \vec x}+h.c.\).
\eea 
The mode functions $u(\eta,\vec k)e^{i\vec k\cdot \vec x}$ and $v^\lambda_\mu (\vec k)e^{i\vec k\cdot \vec x}$ satisfy the equations of motion for scalar and vector fields respectively. The equations of motion are, 
\bea \label{Eq: KG in FRW} (\partial_\mu\partial^\mu+2 \HH(\eta)\partial_\eta+m^2a^2(\eta))\phi=0 \eea
\bea \label{Eq: Maxwell in FRW} \partial^\mu F_{\mu\nu}+m^2a^2(\eta)A_\nu=0.\eea
 To simplify, we make use of the fact that rate at which the universe is expanding ($\HH\equiv \partial_\eta \ln(a(\eta))$) is much slower than the rate at which our fields are oscillating, which is roughly the comoving CMB temperature $T_{CMB}$. In this limit one can show that up to corrections of $\OO(\vec k/\HH)$,  
\bea 
\label{Eq: FRW Ansatz Soln}
u(\vec k, \eta)=\frac{e^{-i\int^\eta d\tilde \eta \omega^c(\eta,\vec k)}}{a(\eta)\sqrt{2\omega^c(\eta)}}\quad \text{and} \quad v^\lambda_\mu(\vec k, \eta)=\frac{\epsilon_\mu^\lambda(\vec k,\eta)e^{-i\int^\eta d\tilde \eta \omega^c(\eta,\vec k)}}{\sqrt{2\omega^c(\eta)}}.
\eea
The mode functions are normalized by making sure they reduce to the familiar flat space mode functions in the flat space limit. Here the $\epsilon_\mu^\lambda$ represent the 3 different polarizations for the vector
\bea 
\label{Eq: Polarizations}
\epsilon_L^\mu=\frac{1}{a(\eta)m}\(|\vec k|,\omega^c(\eta,\vec k)\hat{\vec k}\)\qquad \epsilon^\mu_{1,2}(\vec k)=(0,\vec \epsilon) \where \vec k \cdot \vec \epsilon=0 
\eea
which are the usual transverse and longitudinal polarizations with $m\rar a(\eta) m$. Inserting Eq.~\ref{Eq: FRW Ansatz Soln} into Eq. ~\ref{Eq: FRW Ansatz} gives Eq.~\ref{Eq: FRW Scalar Field} and~\ref{Eq: FRW Vector Field}.

\subsection{The Interaction Potential}
Here we simplify the interaction potentials given in Eq.~\ref{Eq: Interaction Pot Simplified}. We begin as described, by inserting field operators for the incoming photon and outgoing $X$ particle, and a classical field background for the dark matter field, leading to
\bea
\label{Eq: VAxion}
V^{\phi}_I(\eta')=&&-\int \frac{d^3\vec x d^3\vec p d^3\vec p'}{(2\pi)^62\sqrt{\omega^c_\gamma (\eta',\vec p)\omega^c_D(\eta',\vec p')}}\\ &&\frac{\partial_\eta \phi(x)}{f_a}
\sum_{\lambda,\lambda'}  \(a^{\dag,\gamma_D}_{\vec p',\lambda'}a^{\gamma}_{\vec p,\lambda}\vec \epsilon^*_{\lambda'}(\vec p')\cdot \(i\vec p\times\vec \epsilon_{\lambda}(\vec p)\) e^{-ix\cdot ( p-p')}+...\) \, , \nn
\eea
\bea
\label{Eq: VDP}
V^{D}_I(\eta')=&&\int \frac{d^3\vec x d^3\vec p d^3\vec p'}{(2\pi)^62a(\eta')\sqrt{\omega^c_\gamma (\eta',\vec p)\omega^c_\phi(\eta',\vec p')}}\\ &&\frac{\partial_\eta\vec A_D(x)}{f_a}\cdot 
\sum_{\lambda} \(a^{\dag,\phi}_{\vec p'}a^{\gamma}_{\vec p,\lambda} \(i\vec p\times\vec \epsilon_{\lambda}(\vec p)\) e^{-ix\cdot ( p-p')}+...\)\nn \, ,
\eea
where $x\cdot ( p-p')$ is a short hand notation for
\begin{equation}
    x\cdot ( p-p')\equiv \int^{\eta'}d\tilde \eta \(\omega^c_\gamma(\tilde \eta,\vec p)-\omega^c_X(\tilde \eta,\vec p')\)-\vec x\cdot (\vec p-\vec p') \, ,
\end{equation}
and recalling that $V^\phi$ ($V^D$) is the interacting potential when the axion (dark photon) is dark matter.

The terms not explicitly written in Eqs.~\ref{Eq: VAxion} and~\ref{Eq: VDP} represent different combinations of the ladder operators that will be irrelevant for us since we only wish to consider photons as the initial state and outgoing axions/dark photons in the final state. In the regions of parameter space we will be interested in, the dark matter mass will be smaller than the CMB temperature. Since dark matter is also non-relativistic, this means that the momentum transfer from the dark matter, $\vec q\equiv \vec p-\vec p'$, must be small with respect to the photon momentum,
\bea 
\label{Eq: Small Momentum}
|\vec q|\ll m_{DM}\ll T_{CMB}\sim |\vec p| \, ,
\eea
so we can do an expansion in small $|\vec q|$. To lowest order, this means setting $\vec p=\vec p'$ everywhere. However, we must keep $\vec q$ to linear order in the $x\cdot (p-p')$ order since $\vec q\cdot \vec x$ is not necessarily small since we will be integrating over all $\vec x$. 
\bea
\label{Ea: Expanding Exp}
x\cdot ( p-p')\approx \int^{\eta'}d\tilde \eta \(\omega^c_\gamma(\tilde \eta,\vec p)-\omega^c_X(\tilde \eta,\vec p)\)-\vec q\cdot (\vec x-\hat{\vec p}\eta').
\eea
Now, the only $\vec q$ dependence is in the exponent and we can shift variables $d^3\vec p'\rar d^3\vec q$. After integrating over $d^3\vec q$ we get a delta function $\delta^3(\vec x-\tilde{\vec x})$ where $\tilde{\vec x}=\hat {\vec p}\eta$.~\footnote{
Note that, within the approximations we are using, $\tilde x$ is effectively the position of the photon. This is in agreement with the intuition that the transition probability of a photon at a given location depends on the dark matter field at that same location.
} This can then be used to eliminate the $d^3\vec x$ integral. 
\bea
\label{Eq: VAxion 2}
V^{\phi}_I(\eta')=&&-\int \frac{ d^3\vec p }{(2\pi)^3}\frac{\partial_\eta \phi(\eta',\tilde{\vec x}) }{2f_a\sqrt{\omega^c_\gamma (\eta')\omega^c_D(\eta')}}\\ &&\sum_{\lambda,\lambda'} \vec \epsilon^*_{\lambda'}(\vec p)\cdot \(i\vec p\times\vec \epsilon_{\lambda}(\vec p)\)
e^{ i\int^{\eta'}d\tilde \eta \(\omega^c_D(\tilde \eta)-\omega^c_\gamma(\tilde \eta)\)}a^{\dag,\gamma_D}_{\vec p,\lambda'}a^{\gamma}_{\vec p,\lambda}+...\nn
\eea
\bea
\label{Eq: VDP 2}
V^{D}_I(\eta')=&&\int \frac{ d^3\vec p }{(2\pi)^3}\frac{\partial_\eta \vec A_D(\eta',\tilde{\vec x}) }{2f_a a(\eta')\sqrt{\omega^c_\gamma (\eta')\omega^c_D(\eta')}}
\cdot \\ 
&&\sum_{\lambda} \(i\vec p\times\vec \epsilon_{\lambda}(\vec p)\)e^{ i\int^{\eta'}d\tilde \eta \(\omega^c_\phi(\tilde \eta)-\omega^c_\gamma(\tilde \eta)\)}a^{\dag,\phi}_{\vec p}a^{\gamma}_{\vec p,\lambda}+...\;.\nn
\eea
 
 Now, in order to simplify the cross products, we work in the helicity basis for photon polarizations where the following identities hold.
\bea
\label{Eq: Helicity basis ID}
 i\vec p\times \epsilon_\lambda(\vec p)=\lambda|\vec p| \vec \epsilon_\lambda \quad \quad \vec \epsilon_\lambda^*\cdot \vec \epsilon_{\lambda'}=\delta_{\lambda\lambda'},
\eea
where $\lambda=\pm 1$ is the helicity of the photon. This helps simplify the expressions to

\bea
\label{Eq: VAxion 2}
V^{\phi}_I(\eta')=-\int \frac{ d^3\vec p }{(2\pi)^3}\sum_{\lambda}\frac{\lambda|\vec p|\partial_\eta \phi(\eta',\tilde{\vec x}) }{2f_a\sqrt{\omega^c_\gamma (\eta')\omega^c_D(\eta')}}
e^{ i\int^{\eta'}d\tilde \eta \(\omega^c_D(\tilde \eta)-\omega^c_\gamma(\tilde \eta)\)}a^{\dag,\gamma_D}_{\vec p,\lambda}a^{\gamma}_{\vec p,\lambda}+...
\eea
\bea
\label{Eq: VDP 2}
V^{D}_I(\eta')=\int \frac{ d^3\vec p }{(2\pi)^3}\sum_{\lambda}\frac{\lambda|\vec p|\partial_\eta \vec A_D(\eta',\tilde{\vec x})\cdot\vec \epsilon_{\lambda}(\vec p) }{2f_a a(\eta')\sqrt{\omega^c_\gamma (\eta')\omega^c_D(\eta')}}
e^{ i\int^{\eta'}d\tilde \eta \(\omega^c_\phi(\tilde \eta)-\omega^c_\gamma(\tilde \eta)\)}a^{\dag,\phi}_{\vec p}a^{\gamma}_{\vec p,\lambda}+...\;.
\eea
 Next, we can look at $V_I$'s matrix elements with momentum eigenstates. After some simplification, it is easy to see these matrix elements take the form,
\bea
\label{Eq: Matrix Element}
\bra \vec k',(\lambda')|V_I(\eta')|\vec k,\lambda\ket =ia(\eta')\MM^\lambda(\eta',\vec k)(2\pi)^3\delta^3 (\vec k-\vec k')(\delta_{\lambda'\lambda}),
\eea
where, the $(\lambda')$ and $(\delta_{\lambda\lambda'})$ are meant to be included if the final state is a dark photon and $\MM^\lambda(\eta',\vec k)$ are given by Eq.~\ref{Eq: Prob Form A} and~\ref{Eq: Prob Form DP}. After dropping  the factor $(2 \pi)^3\delta^3(\vec k-\vec k')$ due to state normalization, Eq.~\ref{Eq: Matrix Element} can then be inserted into Eq.~\ref{Eq: Transition Amplitude} and squared to yield Eq.~\ref{Eq: Prob Form}-\ref{Eq: Prob Form DP} for the conversion probability.

\subsection{Spatial Averages}
\label{App: Avgs}
In this section, we argue that in all distortions we are averaging the interaction position $\tilde{\vec x}$ over many de Broglie wavelengths of the dark matter field. This can be easily justified given that we are interested in the monopole spectrum, and thus will average distortions over all directions. This effectively means we will be averaging over many de Broglie wavelengths of the dark matter field. For the pre-recombination distortions, we are interested in the averaged conversion rate as a function of redshift, which depend on the conversion probability between photon scatterings. Thus, we average over all possible photon trajectories, and thus $\tilde{\vec x}$ everywhere in space.

For the free streaming case, one can make a more general argument, which shows that even if one is interested in anisotropies, this averaging is justified. First note that the smallest dark matter mass we consider is $10^{-22}$ eV, and so the largest de Broglie wavelengths we must consider are $\lesssim 10$ kpc, which is much smaller than the horizon size today. Because the probability conversion depends on the dark matter density, it is dominated at larger redshifts. This means that for photon conversions happening at similar times by directions separated by $\delta \theta \sim 1$, the distance between the transition points is $\sim H_0^{-1} * a_t \gg 10$ kpc, where $a_t$ is the redshift of the transition.

Finally, let us average the dark photon dark matters polarization $\vec \epsilon(\vec x)$ over $\vec x$ and derive the replacement given in Eq.~\ref{Eq: VDM Polarization Average}. To start notice that when plugging Eq.~\ref{Eq: Prob Form DP} into Eq.~\ref{Eq: Prob Form}, There will be a factor that looks like 
\bea
\label{Eq: Vector Average}
P\propto \frac{1}{2}\sum_\lambda|\vec \epsilon^\lambda(\vec k)\cdot \vec \epsilon(\tilde{\vec x})|^2=\frac{\delta_{ij}-\hat k_i\hat k_j}{2} \epsilon^i(\tilde{\vec x})\epsilon^{*j}(\tilde{\vec x}).
\eea
If we call the $\vec \hat k$ direction the $z$ direction, then this is simply,
\bea
 \frac{1}{2}\sum_\lambda|\vec \epsilon^\lambda(\vec k)\cdot \vec \epsilon(\tilde{\vec x})|^2=\frac{1-|\epsilon_z(\tilde{\vec x})|^2}{2}.
\eea
Now we average over all possible $z$-compontents of the dark photons polarization which gives, 
\bea
 \frac{1}{2}\sum_\lambda|\vec \epsilon^\lambda(\vec k)\cdot \vec \epsilon(\tilde{\vec x})|^2=1/3.
\eea
Thus the effect of averaging is simply to send $\vec \epsilon^\lambda(\vec k)\cdot \vec \epsilon(\tilde{\vec x})\rar 1/\sqrt{3}$ as described in Eq.~\ref{Eq: VDM Polarization Average}.

\section{Free Distortion Computation}
\label{App: Free Dist Comp}
\begin{figure}[h]
    \centering
    \includegraphics[width=0.75\textwidth]{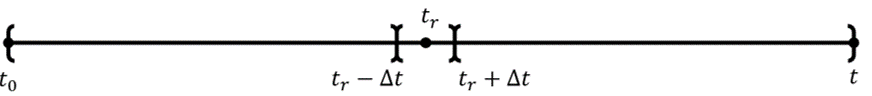}
    \caption{Breaking the time interval between $t$ and $t_0$ into subintervals that either contain or do not contain a resonance. Note that this is easily extended to the case of multiple resonances. The size of the intermediate integral is greatly exaggerated so that it is visible. }
    \label{Fig: Res Intervals}
\end{figure}

In this section we will detail the computation of the integral $L^2$ defined in Eq.~\ref{Eq: L Def} which we showed can be written as 

\bea 
\label{Eq: Lpm w/Omega}
L^2=\bra\bigg|\frac{L_++L_-}{2}\bigg|^2\ket_\beta \where 
L_\pm=\int_{t_0}^t dt'\sqrt{\frac{v_X(t')}{a^3(t')}}e^{i\int^{t'}d\tilde t \Omega_\pm(t')}.
\eea
The strategy to computing this integral is to break it into intervals in which we can use either the fast oscillation limit or the stationary phase approximation. By choosing the boundaries of this regions appropriately, we can piece these intervals together to get the full result. We first identify any resonant times $t_r$ by solving Eq.~\ref{Eq: Resonant Condition} for $t_r$. We can then break up the time interval into sub intervals as shown in Fig.~\ref{Fig: Res Intervals}.

We will choose the endpoints of these intervals $t_r\pm \Delta t$ so that the following 3 conditions are true. 
\begin{enumerate}
    \item The stationary phase approximation should be valid everywhere inside the interval $(t_r-\Delta t,t_r+\Delta t)$ such that we can expand the phase of the exponential to second order 
    \bea
    \label{Eq: second order stationary phase}
    \int^{t'}d\tilde t\, \Omega_\pm(\tilde t)\approx \int^{t_r}d\tilde t\, \Omega_\pm(\tilde t)+\dot \Omega_\pm(t_r)\frac{\Delta t^2}{2}.
    \eea
    In order for this approximation to be valid, we must be able to ignore the third order term. This means we need
    \bea
    \label{Eq: Stationary Validity}
    \Delta t\ll \frac{\dot \Omega_\pm(t_r)}{\ddot\Omega_\pm(t_r)}\sim H^{-1}(t_r).
    \eea
    \item In order to match regions where the stationary phase approximation is valid to regions where the fast oscillation condition is valid, we want both the fast oscillation condition, Eq.~\ref{Eq: Fast Cond}, and the stationary phase approximation to be valid at the endpoints $t_r\pm \Delta t$. Given that the stationary phase approximation is valid, we can write 
    \bea \Omega(t_r\pm \Delta t)\approx \pm\dot \Omega(t_r)\Delta t.
    \eea
    Then since we are near resonance $\dot \Omega(t_r)\sim m_{DM} H$ and we find that Eq.~\ref{Eq: Fast Cond} requires 
    \bea
    \label{Eq: Fast Endpt Cond}
    1\ll m_{DM}\Delta t.
    \eea
    \item Finally for computational ease, in the stationary phase integrals, we want to be able to take the limit $\Delta t\rar \infty$. More precisely this will require that $\Delta t$ is much larger than the spread of the Gaussian integrand, $\sqrt{\dot\Omega_\pm} \sim \sqrt{m_{DM}H}$. We are then able to take the $\Delta t\rar \infty$ limit as long as,  
    \bea
    \label{Eq: t goes to infty cond}
    1\ll \sqrt{H(t_r)m_{DM}}\Delta t.
    \eea
\end{enumerate}
It is easy to see that all 3 conditions are satisfied if
\bea
\label{Eq: Endpoint condition}
1\gg H(t_r)\Delta t\gg \sqrt{\frac{H(t_r)}{m_{DM}}}.
\eea
Using $m_{DM}\geq 10^{-22}$ eV and $H\leq 10^{-29} $ eV, this translates to $1\gg H(t_r)\Delta t\gg 10^{-4}$ which is easily satisfied. This shows that we are able to choose endpoints that satisfy all of the 3 conditions. Then from the first and second condition, as shown in Fig.~\ref{Fig: Res Intervals}, the total $L_\pm$ can be broken into a series of alternating resonant and fast contributions. However, as we will see, the resonant pieces will always dominate over the fast contributions. We will find that these terms take the form
\bea 
\label{Eq: Lpm conditional}
L_\pm =\begin{cases} e^{\mp i \beta(t_0)}L_\pm^{fast}(t_0)& \text{if there are no resonances}\\
\sum_{t_r} e^{\mp i \beta(t_r)}L_\pm^{res}(t_r)& \text{if there are resonances}.
\end{cases}
\eea
Where $L_\pm^{fast}(t_0)$ and $L_\pm^{res}(t_r)$ do not depend on the dark matter phase $\beta$. This allows us to square and average over the phase which simply eliminates any terms that get a non-trivial phase. This eliminates not only cross terms between the $L_+$ and $L_-$ pieces, but any cross terms between different resonances. The end result is 
\bea
\label{Eq: Lsq final}
L^2=\frac{|L_+|^2+|L_-|^2}{4}\where |L_\pm|^2=\begin{cases} |L_\pm^{fast}(t_0)|^2& \text{if there are no resonances}\\
\sum_{t_r} |L_\pm^{res}(t_r)|^2& \text{if there are resonances}.
\end{cases}
\eea
Note that $L_+$ and $L_-$ have different sets of resonant times. Finally, we must compute the fast and resonant integrals to find $|L_\pm^{fast}|$ and $|L_\pm^{res}|$.

\subsection{Fast limit}
In the limit of fast oscillations we assume $\Omega_\pm\gg H$ for the entire integral. Then we can rewrite Eq.~\ref{Eq: Lpm w/Omega} as
\bea
L_\pm=\int_{t_0}^tdt'\sqrt{\frac{v_X(t')}{a^3(t')}}\frac{1}{i\Omega_\pm(t')}\frac{d}{dt}e^{i\int^{t'} d\tilde t\,\Omega_\pm(\tilde t)}.
\eea 
This can then be integrated by parts
\bea
L_\pm=\sqrt{\frac{v_X(t')}{a^3(t')}}\frac{1}{i\Omega_\pm(t')}e^{i\int^{t'} d\tilde t\,\Omega_\pm(\tilde t)}\bigg|_{t_0}^t-\int_{t_0}^tdt' e^{i\int^{t'} d\tilde t\,\Omega_\pm(\tilde t)}\frac{d}{dt}
\(\sqrt{\frac{v_X(t')}{a^3(t')}}\frac{1}{i\Omega_\pm(t')}\).
\eea
Since the time derivative in the second term is hitting quantites that change on the Hubble scale, this second term represents an $\OO(H/\Omega_\pm)$ correction and can be ignored. In the first term, the piece evaluated at $t$ can be ignored due to the scale factor in the denominator. Finally, since $\dot\beta\ll m_{DM}$ we can ignore this term in the denominator, leaving
\bea
L_\pm^{fast}=e^{\mp i\beta(t_0)}\sqrt{\frac{v_X(t_0)}{a^3(t_0)}}\frac{-i}{\Delta\omega(t_0)\pm m_{DM}}e^{i\int^{t_0} d\tilde t (\Delta\omega(\tilde t)\pm m_{DM})}.
\eea
We see we get the exact $\beta$ dependence predicted in Eq.~\ref{Eq: Lpm conditional}. We find then
\bea
\label{Eq: Lfast}
|L_\pm^{fast}|=\sqrt{\frac{v_X(t_0)}{a^3(t_0)}}\frac{1}{|\Delta\omega(t_0)\pm m_{DM}|}.
\eea

\subsection{Resonant Limit}
Now we look at the resonant integral.
\bea 
\label{Eq: Stationary Phase}
L_\pm^{res}=\int_{t_r-\Delta t}^{t_r+\Delta t}dt'\sqrt{\frac{v_X(t')}{a^3(t')}}e^{i\int^{t'}d\tilde t\Omega_\pm(t')}.
\eea
Expanding the integrand to leading order about the resonance time $t_r$ gives 
\bea 
\label{Eq: Stationary Phase}
L_\pm^{res}=\sqrt{\frac{v_X(t_r)}{a^3(t_r)}}e^{i\int^{t_r}d\tilde t\,\Omega_\pm(t')}\int_{t_r-\Delta t}^{t_r+\Delta t}dt'e^{i\frac{\dot\Omega_\pm(t_r)}{2}(t'-t_r)^2}.
\eea
This is simply a gaussian integral and can be easily computed in the limit $\Delta t \rar \pm \infty$. 
\bea 
\label{Eq: Lpm Res}
L_\pm^{res}=e^{\mp i\beta(t_r)}\sqrt{\frac{2\pi}{i\dot \Omega_\pm(t_r)}\frac{v_X(t_r)}{a^3(t_r)}}e^{i\int^{t_r}d\tilde t\,(\Delta\omega (\tilde t)\pm m_{DM})}.
\eea
To leading order, $\dot \Omega_\pm(t_r)=\dot{\Delta\omega}(t_r)$. The only remaining $\beta$ dependence is in the phase. We then find 
\bea 
\label{Eq: Mag Lpm Res}
|L_\pm^{res}|=\sqrt{\frac{2\pi}{|\dot {\Delta\omega}(t_r)|}\frac{v_X(t_r)}{a^3(t_r)}}.
\eea
Combining this result with Eq.~\ref{Eq: Lfast} and Eq.~\ref{Eq: Lsq final} and inserting into Eq.~\ref{Eq: Probabilities} yields~\ref{Eq: Free Dist Final}.

\section{Pre-recombination Distortion Computation}
Here we give some of the details of the pre-recombination distortion. First, we describe how to use the probabilities given in Eq.~\ref{Eq: Probabilities} to compute the rate of photon conversion in Eq.~\ref{Eq: Rate final}. Second we describe the Green's function method for using this rate to compute the distortions in different eras. 

\subsection{Photon Conversion rate}
\label{App: Photon Rate}

Here we will give the details of computing the conversion rate for photons into particle $X$ given in Eq.~\ref{Eq: Rate averaged}. Because the photons are not free streaming, our integral is over a small time interval, $\tau$, with respect to the expansion rate $H$. Therefore, quantities which depend on time through the expansion of the universe are approximately constant. We then can write,
\bea 
\label{Eq: P Trapped Start}
P(\vec k,t_0,t_0+\tau)=\frac{\rho_{DM}^0v_X(a(t_0))}{8f_a^2a^3(t_0)}\bigg|\ell_++\ell_-\bigg|^2 \where \ell_\pm=\int_0^{\tau} dt e^{i\Omega_\pm(a(t_0)) (t+t_0)}.
\eea
This can be easily integrated. 
\bea 
\ell_\pm=2e^{i \Omega_\pm (t_0+\tau/2 )}\frac{\sin(\Omega_\pm\tau/2)}{\Omega_\pm}.
\eea
Now we will square this and average over the dark matter phase $\beta$ as discussed in Appendix~\ref{App: Avgs}. From the $\Omega_\pm$ in the exponential we get a factor of $e^{\mp i\beta(t_0+\tau/2)}$. Everywhere else we can ignore $\beta$ because it is sub-leading. Then just as for the free streaming distortion, the phase averaging eliminates the cross terms between $\ell_+$ and $\ell_-$. This gives,
\bea 
\label{Eq: P Trapped Start}
\bra P(\vec k,t_0,t_0+\tau)\ket_\beta &&=\frac{\rho_{DM}^0v_X(a)}{2f_a^2a^3}\\ &&\(\frac{\sin^2((\Delta \omega+m_{DM})\tau/2)}{(\Delta \omega+m_{DM})^2}+\frac{\sin^2((\Delta \omega-m_{DM})\tau/2)}{(\Delta \omega-m_{DM})^2}\).\nn
\eea
Finally, one can plug this into Eq.~\ref{Eq: Rate averaged} and evaluate that integral analytically to get Eq.~\ref{Eq: Rate final}. 

\subsection{Green's Function Method}
\label{App: Green's Function}
Here we present the Green's function for the different eras used in Eq.~\ref{Eq: Green's function} and explain how to compute the parameters $\bar \mu$, $\bar \mu_t$, $y_t$ and $y$ in Eq~\ref{Eq: Total Distortion}. The Green's functions used are taken from Ref.~\cite{Chluba:2015hma} for the $\mu$ and $y$ era and from Ref.~\cite{Chluba:2013vsa} for the $\mu-y$ transition era and modified by absorbing and moving a few factors to fit with the definition in Eq~\ref{Eq: Green's function}. To start let us define the temperature shift function $\TT(x)$, the $\mu$ distortion shape $M(x)$, and the $y$ distortion shape $Y(x)$

\bea 
\label{Eq: Distortion Shapes} \TT(x)=\frac{xe^x}{(e^x-1)} \qquad M(x)=\TT(x)\(\alpha_\mu-\frac{1}{x}\)\qquad Y(x)=\TT(x)\(x\frac{e^x+1}{e^x-1}-4\), 
\eea
where $\alpha_\mu=\zeta(2)/3\zeta(3)\approx0.456$\\\\

\noindent \textbf{$\mu$ Era}\\
For the $\mu$ era, the Green's function takes the form 
\bea
\label{Eq: Mu Era Green's}
G_\mu(x,x',a)=1.4\(1-\frac{x_0}{x'}\)\bar\rho(x')J^*(a)M(x)
\eea
Where $x_0=\frac{4\zeta(3)}{\zeta(2)}\approx 3.6$, and $\bar\rho(x')=\frac{15}{\pi^4}\frac{x^3}{e^x-1}$ is the unit-normalized blackbody energy spectrum. $J^*(a)$ is called the visibility function and captures how inefficient bremsstrahlung and double Compton scattering are at changing the number density. $J^*(a)$ goes to 0 for $a\ll 10^{-6}$ and quickly goes to zero at early times. Its analytic form can be found in Eq. 13 of Ref.~\cite{Chluba:2015hma}. Note that the $M(x)$ factors out completely and we can write the distortion as 
\bea\label{Eq: Mu param}
\delta(x)=\bar\mu M(x) \where \bar \mu =1.4\int dx'\frac{da}{aH(a)}J^*(a)  \(1-\frac{x_0}{x'}\)\bar\rho(x')\Gamma_{\gamma \rar X}(x',a),
\eea
where the $a$ integral runs from $0$ to $a=3.3\cdot 10^{-6}$.\\\\

\noindent\textbf{$\mu$-y Transition Era}\\
The Green's function for the transition era is similar to that of the $\mu$ era with the addition of a $y$-distortion piece.
\bea 
\label{Eq: Transition Era Green's}
G_{t}(x,x',a)=\(1-\frac{x_0}{x'}\)\bar\rho(x')\( 1.4 J^*(a)J_\mu (a)M(x)+\frac{1}{4}J_y(a)Y(x)\).
\eea
The additional factors $J_\mu$ and $J_y$ smoothly transition the Green's function from having mostly $\mu$ distortion at early times in the era to mostly $y$- distortions late in the era. Their analytical form can be found in Eq.~5 of Ref.~\cite{Chluba:2013vsa}. Much like in the $\mu$ era we can write this distortion as 
\bea\label{Eq: Trans distortion}
\delta(x)=\bar\mu_tM(x)+y_tY(x),
\eea
where 
\bea
\label{Eq: Mu t def}
\bar\mu_t=1.4\int dx'\frac{da}{aH(a)}J^*(a)J_\mu (a)\(1-\frac{x_0}{x'}\)\bar\rho(x')\Gamma_{\gamma \rar X}(x',a)
\eea 
\bea
\label{Eq: y t def}
y_t=\frac{1}{4}\int dx'\frac{da}{aH(a)}J_y (a)\(1-\frac{x_0}{x'}\)\bar\rho(x')\Gamma_{\gamma \rar X}(x',a)
\eea 
and the $a$ integral runs over the whole transition era from $a=3.3\cdot 10^{-6}$ to $a=2\cdot 10^{-5}$.\\\\
\noindent\textbf{$y$ Era}\\
The Green's function for the $y$ era contains two terms. One to describe the Doppler smearing of the removed photons, and another to describe the pure $y$-distortion. 
\bea
\label{Eq: G y era} \nonumber
G_y(x,x',a)=\bar\rho(x')\(\(1-\frac{e^{(\alpha+\beta)y_\gamma(a)}}{1+x'y_\gamma(a)}\)\cdot \frac{Y(x)}{4}+\frac{e^{-\(\ln\(x(1+x'y_\gamma(a))/x'\)-\alpha y_\gamma(a)\)^2/4\beta y_\gamma}}{x'\sqrt{4\pi \beta y_\gamma(a)}}\),
\eea
where $y_\gamma(a)$ is the Compton $y$ parameter defined as 
\bea
\label{Eq: compton y param}
y_\gamma(a)\equiv \int_a^1\frac{da'}{a'H(a')}\frac{T(a')}{m_e\lambda_\gamma(a')}
\eea
and $\alpha$ and $\beta$ are defined as 
\bea
\label{Eq: alpha beta def}
\alpha =\frac{3-2f(x')}{\sqrt{1+x' y_\gamma(a)}}\quad \beta=\frac{1}{1+x' y_\gamma(a)(1-f(x'))}
\where f(x')=e^{-x'}\(1+\frac{{x'}^2}{2}\).
\eea 
By definition $y_\gamma$ quantifies the efficiency of Compton scattering to redistribute energy and is thus $\OO(1)$ at the start of the $y$ era and quickly falls to be much less than 1. We can write the distortion as 
\bea\label{Eq: y era dist}
\delta(x)=y Y(x)+\delta_{Doppler}(x),
\eea 
where 
\bea 
\label{Eq: y def}
y=\frac{1}{4}\int dx'\frac{da}{aH(a)} \bar\rho(x')\(1-\frac{e^{(\alpha+\beta)y_\gamma(a)}}{1+x'y_\gamma(a)}\)\Gamma_{\gamma \rar X}(x',a)
\eea
\bea 
\label{Eq: Doppler def}
\delta_{Doppler}(x)=\int dx'\frac{da}{aH(a)}\frac{e^{-\(\ln\(x(1+x'y_\gamma(a))/x'\)-\alpha y_\gamma(a)\)^2/4\beta y_\gamma}}{x'\sqrt{4\pi \beta y_\gamma(a)}} \Gamma_{\gamma \rar X}(x',a).
\eea
Here the $a$ integral runs from $a=2\cdot 10^{-5}$ to the time of recombination at $a_*$.

\bibliographystyle{JHEP}
\bibliography{aAADCMBConstraints}{}

\end{document}